\documentclass[usenatbib]{mn2e}

\usepackage{amsmath, mathrsfs, amssymb}
\usepackage{graphicx}
\usepackage{morefloats}
\usepackage{deluxetable}
\usepackage{multicol}
\usepackage{color}
\usepackage{bm}
\usepackage{yhmath}
\usepackage{enumitem}
\usepackage{caption}

\newcommand{\del}{\partial}
\newcommand{\p}{\prime}
\newcommand{\mach}{\mathscr{M}}
\newcommand{\Alfven}{Alfv\'en }

\title[Energetics and dissipation in MHD turbulence]{Quantifying energetics and dissipation in magnetohydrodynamic turbulence}

\author[Salvesen et al.]{Greg~Salvesen$^{1,2}$\thanks{E-mail: salvesen@colorado.edu}\thanks{National Science Foundation Graduate Fellow.}, Kris~Beckwith$^{1,3}$, Jacob~B.~Simon$^{1,4}$\thanks{Sagan Fellow.}, Sean~M.~O'Neill$^{1,5}$, \& \newauthor
Mitchell~C.~Begelman$^{1,2}$ \\
$^{1}${JILA, University of Colorado and National Institute of Standards and Technology, 440 UCB, Boulder, CO 80309-0440, USA.} \\
$^{2}${Department of Astrophysical and Planetary Sciences, University of Colorado, 391 UCB, Boulder, CO 80309-0391, USA.} \\
$^{3}${Tech-X Corporation, 5621 Arapahoe Ave. Suite A, Boulder, CO 80303, USA.} \\
$^{4}${Southwest Research Institute, 1050 Walnut St. \# 300, Boulder, CO 80302, USA.} \\
$^{5}${Department of Physics, Pacific Lutheran University, 1010 112$^{\rm nd}$ St. S., Tacoma, WA 98447 USA.}}

\begin{document}
\label{firstpage}
\maketitle

\begin{abstract}
We perform a suite of two- and three-dimensional magnetohydrodynamic (MHD) simulations with the \texttt{Athena} code of the non-driven Kelvin-Helmholtz instability in the subsonic, weak magnetic field limit.  Focusing the analysis on the non-linear turbulent regime, we quantify energy transfer on a scale-by-scale basis and identify the physical mechanisms responsible for energy exchange by developing the diagnostic known as spectral energy transfer function analysis.  At late times when the fluid is in a state of MHD turbulence, magnetic tension mediates the dominant mode of energy injection into the magnetic reservoir, whereby turbulent fluid motions twist and stretch the magnetic field lines.  This generated magnetic energy turbulently cascades to smaller scales, while being exchanged backwards and forwards with the kinetic energy reservoir, until finally being dissipated.  Incorporating explicit dissipation pushes the dissipation scale to larger scales than if the dissipation were entirely numerical.  For scales larger than the dissipation scale, we show that the physics of energy transfer in decaying MHD turbulence is robust to numerical effects.
\end{abstract}


\section{Introduction}
\label{sec:intro}
The nature of magnetized gas in astrophysical systems is a long standing problem.  Linear analyses of various fluid configurations demonstrate that instabilities expected to be relevant in astrophysical contexts, such as the magnetorotational instability \citep[MRI;][]{BalbusHawley1991} or the Kelvin-Helmholtz instability \citep[KHI; e.g.,][]{Chandrasekhar1961}, can amplify the magnetic field and generate turbulence.  However, the subsequent non-linear evolution into magnetohydrodynamic (MHD) turbulence is only accessible by appealing to numerical simulations.

Spectral energy transfer function analysis, first introduced by \citet{Kraichnan1967}, is a powerful diagnostic for quantifying energetics and dissipation in MHD turbulence.  This diagnostic is able to determine precisely how energy is transferred across spatial scales as a function of both energy type (e.g., kinetic, magnetic, internal) and mediating force (e.g., compression, advection, magnetic tension, magnetic pressure).  In addition to probing the energetics of MHD turbulence, transfer function analysis allows for the scale-by-scale characterization of physical and numerical dissipation.  Therefore, transfer function analysis goes beyond the standard power spectrum diagnostic, which only provides information about the distribution of energy across spatial scales and says nothing about either the energy transfer mechanism or the scales on which energy exchange occurs.

Quantifying MHD turbulence with transfer function analysis experienced a recent revival with the advent of high performance numerical simulations.  Transfer function analysis has far-reaching astrophysical applications, including the turbulent solar dynamo \citep{PietarilaGraham2010}, accretion disc turbulence arising from the MRI \citep{FromangPapaloizou2007a, FromangPapaloizou2007b, SimonHawleyBeckwith2009, SimonHawley2009, Davis2010}, and ``mesoscale'' magnetic structures that arise in local studies of accretion discs with large spatial domains \citep{Simon2012}.  The transfer function diagnostic also provides a scale-by-scale look into the properties of energy dissipation distributions.  For instance, applying a transfer function analysis to accretion disc atmospheric and coronal structure could reveal new insights into accretion power dissipation profiles, which have important consequences for the emergent spectra from black hole systems \citep[e.g.,][]{SvenssonZdziarski1994, Merloni2000, Turner2004, Hirose2006, Blaes2006, Salvesen2013}.

The aim of this work is to explore transfer function analysis in detail and further develop this diagnostic for MHD turbulence.  Therefore, we seek a well-understood test problem that can generate an MHD turbulent state from generic initial conditions, but is free from potential complications such as strong magnetic field effects, supersonic motions, and forcing of the turbulence.  Motivated by these criteria, we elect to study the KHI, which is a well-posed linear instability commonly used in code testing \citep{McNallyLyraPassy2012}.

Numerical methods are essential for understanding the non-linear development of the KHI.  In two spatial dimensions (2D), hydrodynamics simulations of shearing flows by \citet{NormanHardee1988} and \citet{Bodoetal1994, Bodoetal1995} provided some of the first glimpses into the non-linear evolution of the KHI, followed by extensions into 2D/axisymmetric MHD by \citet{Franketal1996}, \citet{Jonesetal1997}, \citet{Jeongetal2000}, \citet{StoneHardee2000}, and \citet{Palottietal2008}.  Full three-dimensional (3D) numerical explorations of the KHI were conducted by \citet{NormanBalsara1993} and \citet{BassettWoodward1995} in the hydrodynamic limit and \citet{Hardeeetal1997}, \citet{Ryuetal2000}, and \citet{HardeeRosen2002} in full MHD.  More recently, \citet{Martietal2004}, \citet{Peruchoetal2004a, Peruchoetal2004b}, \citet{Peruchoetal2006}, and \citet{RadiceRezzolla2012} explored the KHI in relativistic hydrodynamics, while \citet{ZhangMacFadyenWang2009} and \citet{BeckwithStone2011} discussed KHI development in relativistic MHD.

The majority of these numerical studies analyzed KHI development through the measurement of instability growth rates, saturation behaviours, and/or morphological consequences of instability.  In this study, we provide a novel look at the development of the KHI by employing spectral energy transfer function analysis.  This approach provides us with insight into the details of the KHI physics that are not otherwise accessible and allows us to determine how integrated flow properties and morphology reflect the scale-dependent processes we identify.  Additionally, we discuss how computational issues, such as numerical convergence and the effects of domain size can be understood and evaluated in terms of KHI development.  We will also explore numerical {\it versus} physical dissipation behaviours by comparing simulations of decaying MHD turbulence with and without dissipation in the same spirit as done previously for simulations of decaying hydrodynamic turbulence \citep[e.g.,][]{Sytine2000}.

While the KHI has important applications to subsonic, transonic, supersonic, and relativistic astrophysics, we focus here on understanding the non-linear development and spectral structure of the KHI in the subsonic, weakly magnetized limit.  The motivations for this choice are both simplicity and applicability.  We wish to apply comprehensive analysis tools --- particularly transfer function analysis --- to study the development of the KHI for a simplified case without the complications of additional physics like shock formation or the exchange between different fluid instabilities such as the family of current-driven instabilities \citep[CDI;][]{Begelman1998}.  Particular attention is given to properly constructing an initial setup for the simulations and providing convincing evidence that the late-stage development is physical, rather than numerical, in origin.  We aim for our numerical study of the KHI to be relevant and extendable to a broad range of astrophysical applications, such as the interplay between the KHI and CDI in jets, the nature of MHD turbulence arising from the KHI, and dissipation profiles in accretion discs.

We organize this paper as follows.  \S \ref{sec:numsims} provides descriptions of the \texttt{Athena} MHD code and the KHI problem setup.  The methodology behind the spectral energy transfer function analysis we adopt is given in \S \ref{sec:specanal}.  In \S \ref{sec:converge}, we discuss the convergence of 3D KHI simulations, along with the inadequacy of 2D simulations.  We next describe in \S \ref{sec:devevo} the evolution of the KHI simulations with a focus on the late-stage turbulent decay and the importance of energy transfer involving the magnetic energy reservoir.  The inclusion of physical dissipation is explored in \S \ref{sec:dissip}.  Finally, \S \ref{sec:sumdisc} presents a summary and discussion of this work, followed by our conclusions in \S \ref{sec:conc}.

\section{Numerical Details}
\label{sec:numsims}
We solve the equations of MHD using the \texttt{Athena} code \citep{Stone2008}, a second-order accurate Godunov flux-conservative code\footnote{The \texttt{Athena} code and a repository of test problems are available online at https://trac.princeton.edu/Athena/.}. \texttt{Athena} is an Eulerian code that solves the equations of compressible, adiabatic MHD in conservative form,
\begin{align}
\frac{\del \rho}{\del t} =& - \nabla \cdot \left( \rho \mathbf{v} \right) \label{eq:continuity}  \\
\frac{\del \left( \rho \mathbf{v} \right)}{\del t} =& - \nabla \cdot \left[ \rho \mathbf{v} \mathbf{v} - \mathbf{B} \mathbf{B} + \left( P + \frac{1}{2} B^{2} \right) \mathbf{I} - \bm{\tau} \right] \label{eq:momentum} \\
\frac{\del E}{\del t} =& - \nabla \cdot \left[ \left( E + P + \frac{1}{2} B^{2} \right) \mathbf{v} - \mathbf{B}\left( \mathbf{B} \cdot \mathbf{v} \right) \right] \label{eq:energy} \\
\frac{\del \mathbf{B}}{\del t} =&~ \nabla \times \left( \mathbf{v} \times \mathbf{B} \right) + \eta \nabla^{2} \mathbf{B} \label{eq:induction}.
\end{align}
The notation is of familiar form, where $\rho$ is the density, $\mathbf{v}$ is the fluid velocity, $P$ is the gas pressure, $\mathbf{B}$ is the magnetic field, and $E$ is the total energy density defined by,
\begin{equation}
E = \epsilon + \frac{1}{2} \rho v^{2} + \frac{1}{2} B^{2}, \label{eq:Etot}
\end{equation}
where $\epsilon = P / \left( \gamma - 1 \right)$ is the internal energy density for an ideal gas and $\gamma$ is the adiabatic index.  $\mathbf{I}$ is the identity matrix operating on the total pressure, $P + B^{2}/2$.  In the adopted notation, the magnetic field absorbs a factor of $\sqrt{\mu / 4 \pi}$, where $\mu = 1$ is the assumed magnetic permeability.  The MHD equations \ref{eq:continuity}--\ref{eq:induction} are conservation equations describing, in order, the conservation of mass, momentum, total energy, and magnetic flux.  Equations \ref{eq:continuity}--\ref{eq:energy} have the generic form of  any conservation equation, where the time derivative of a conserved quantity is equated to the divergence of a flux, in the absence of any source/sink terms.  

Viscosity enters the momentum equation through the stress tensor,
\begin{equation}
\bm{\tau} = \tau_{ij} = 2 \nu \rho \left[ e_{ij} - \frac{1}{3} \left( \nabla \cdot \mathbf{v} \right) \delta_{ij} \right],
\end{equation}
where the fluid is assumed to be isotropic, $\nu$ is the kinematic viscosity, $\delta_{ij}$ is the Kronecker delta function, and the strain rate tensor is $e_{ij} = \frac{1}{2} \left[ \left( \nabla \mathbf{v} \right) + \left( \nabla \mathbf{v} \right)^{\rm T} \right]$.  Explicit dissipation enters the induction equation through the Ohmic dissipation term, $\eta \nabla^{2} \mathbf{B}$, where $\eta$ is the resistivity.  In our treatment of ideal MHD, we neglect dissipation terms such as viscosity (i.e., $\nu = 0$), resistivity (i.e., $\eta = 0$), and conduction.  We investigate the addition of explicit dissipation terms, following the implementation of \citet{SimonHawleyBeckwith2009}, and their affect on the KHI evolution in \S \ref{sec:dissip}.

\citet{GardinerStone2005,GardinerStone2008} describe the basic algorithms implemented in \texttt{Athena} with further details (implementation and multi-dimensional tests) given in \citet{Stone2008}. Specifically, we use the dimensionally unsplit Corner Transport Upwind (CTU) integrator described by \cite{Stone2008} combined with the constrained transport (CT) method of \citet{EvansHawley1988} to maintain the divergence-less nature of the magnetic field in multi-dimensions. \texttt{Athena} implements a variety of options for spatial reconstruction and solution of the Riemann problem. In this work, we use third-order spatial reconstruction in characteristic variables and the HLLC/HLLD Riemann solvers for hydrodynamic/MHD simulations.  We avoid choosing the HLLE solver due to its highly diffusive behaviour (for further information, see Appendix \ref{append:riemann}). In this work, we make extensive use of the conservation properties of \texttt{Athena} to examine exchange of energy between kinetic, magnetic, and internal energy reservoirs.

\subsection{Problem Setup}
\label{sec:setup_sim}
\begin{figure}
  \includegraphics[scale=0.25, trim=0 75 -200 100, clip=true]{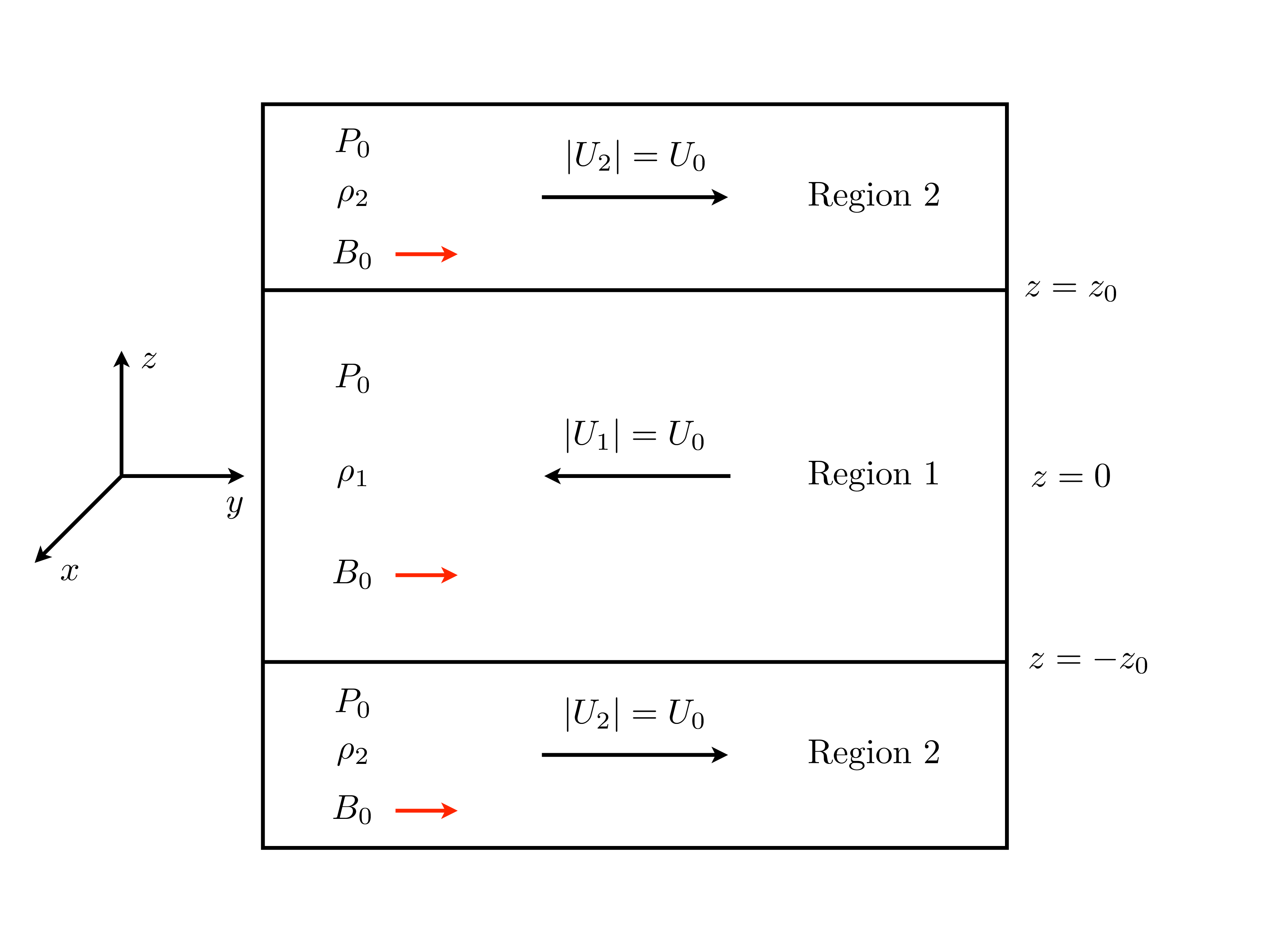}
  \caption{Schematic of the KHI problem setup for the \texttt{Athena} simulations.  Each side of the computational box has length $L$, with the origin at the center, $(x,y,z) = (0,0,0)$.  Periodic boundary conditions are adopted in all directions.  Counter-streaming flows are initiated in the $y$-direction, each with speed $U_{0}$ in the laboratory frame.  A uniform pressure, $P_{0}$, fills the box and a uniform magnetic field of strength $B_{0}$ in the $y$-direction.  The fluid densities are $\rho_{1}$ and $\rho_{2}$ in Region 1 and Region 2, respectively.  Region 1 is bounded in the $z$-direction by $-z_{0} < z < z_{0}$ and Region 2 corresponds to $z > |z_{0}|$, where the shear interfaces are located at $\pm z_{0}$.  Although not represented in the schematic, the density and velocity profiles across each shear layer are matched continuously by a hyperbolic tangent function; thus, permitting the interfaces to be well-resolved.}
  \label{fig:setup_sim}
\end{figure}

The initial problem setup for numerical simulations of the KHI with \texttt{Athena} is shown schematically in Figure \ref{fig:setup_sim}.  We consider a 3D Cartesian grid centered on $\left( x, y, z \right) = \left( 0, 0, 0 \right)$ with dimensions $L_{x} = L_{y} = L_{z} = L = 1$ and periodic boundary conditions enforced in all directions.  Counter-streaming flows are set up along the $y$-direction according to the velocity profile,
\begin{equation}
  v_{y}(z) = \left \{
    \begin{array}{lr}
      U_{0} {\rm tanh} \left( \frac{\left|z\right| - z_{0}}{a}\right), & \left| z \right| \ge z_{0} \\
      -U_{0} {\rm tanh} \left( \frac{z_{0} - \left|z\right|}{a}\right), & \left| z \right| < z_{0}
    \end{array}
  \right.
\end{equation}
where $2U_{0} = 1$ is the magnitude of the relative shear velocity, $z_{0} = 0.25 L$ specifies the location of the shear interfaces, and $a = 0.01 L$ is a parameter describing the thickness of the shear layer.  A continuous velocity profile is constructed across the shear layers, rather than a discontinuous interface, to ensure that truncation error resulting from numerical diffusion of unresolved modes (i.e., short wavelength, large wavenumber) does not dominate the solutions (see Appendix \ref{append:jump}).  The linear growth rate of the KHI is proportional to the wavenumber; thus, an under-resolved shear layer will evolve unphysically into the non-linear regime.  The hyperbolic tangent profile we adopt provides a sharp, yet smooth, transition while also introducing the length scale, $a$, to an otherwise scale-free problem.  For a given grid resolution, $N = N_{x} = N_{y} = N_{z}$, the shear layer is resolved by $4aN/L$ grid zones, which amounts to $\sim 20$ resolution elements across the interface for the $N = 512$ 3D MHD simulation, which is the fiducial run for the majority of the analysis.  The wavenumber corresponding to the full width of the shear layer is $k_{\rm sh} = 2 \pi / (4a) \simeq 157$; therefore, to resolve modes that grow on the same spatial scale of the shear layer or smaller, the simulation resolution must be adequate, such that the Nyquist wavenumber, $k_{\rm Ny} = (N / 2) (2 \pi / L)$, exceeds $k_{\rm sh}$.

The initial density profile is described by,
\begin{equation}
\rho(z) = \frac{1}{2} \left( \frac{\rho_{1}}{\rho_{2}} - 1 \right) \left| {\rm tanh} \left( \frac{\left|z\right| - z_{0}}{a}\right) - 1\right| + 1,
\end{equation}
where $\rho_{1} = 2$ is the density of the central fluid slab and $\rho_{2} = 1$ is the density of the surrounding fluid.  The contact discontinuity is smeared by the same hyperbolic tangent function applied to the velocity profile to ensure a resolved solution.  Initially, the fluid slabs are in gas pressure equilibrium with $P_{0} = 1$ and adiabatic index, $\gamma = 5/3$.  A uniform magnetic field, $\mathbf{B_{0}} = B_{0} \widehat{\mathbf{y}}$, is aligned parallel to the shear flow with initial strength, $B_{0} = 0.02$, corresponding to the weakly non-linear regime, meaning that the magnetic field is weak and the flow is not linearly stable.  In this regime, the instability is essentially hydrodynamic early on, then enters the non-linear regime where secondary instabilities break up large-scale structures and magnetic energy is amplified due to twisting/stretching of magnetic field.  After saturation, the flow enters a state of decaying MHD turbulence \citep[for a discussion of different stability regimes of the magnetized KHI, see][]{Ryuetal2000, BatyKeppens2002}.

In order to provoke the onset of the KHI, at time $t= 0$ we impose a small-amplitude, single mode perturbation to the $z$-component of velocity of the form,
\begin{equation}
\mathbf{v}_{z}^{\p} \left( x, y, z \right) = v_{z}^{0} {\rm sin} \left( k_{x} x \right) {\rm sin} \left( k_{y} y \right) e^{- (z + z_{0})^{2} / \sigma^{2}} \widehat{\mathbf{z}}, \label{eq:pert}
\end{equation}
where $v_{z}^{0} = 0.01 U_{0}$, $k_{x} = k_{y} = 2 \pi / L$, and $\sigma = 0.1 L$ describes the decaying behaviour of the perturbation along the $z$-direction.  A full perturbation wavelength fits within the $x$ and $y$ computational box boundaries.  Modes with wavelengths larger than the box scale, $L$, are not captured.

Table \ref{tab:initial_params} summarizes the set of initial parameters corresponding to each region defined in Figure \ref{fig:setup_sim}.  All simulations used the foregoing setup and parameter choices, unless specified otherwise.  A parameter survey is beyond our scope and does not address our motivating intention to study in detail the development and energetics of the KHI starting from a properly constructed initial configuration.  Table \ref{tab:simlist} lists the suite of simulations presented in this work.

\begin{table}
\addtolength{\tabcolsep}{-1.5pt}
\centering
\begin{tabular}{c c c c c c c c c c}
\hline
\hline
Region & $v_{y}$ & $P_{0}$ & $\rho$ & $c_{s}$ & $\mach$ & $B_{0}$ & $\beta_{0}$ & $v_{A}$ & $\mach_{A}$ \\
\hline
1 & --0.5 & 1 & 2  & 0.91 & 0.39 & 0.02 & 5000 & 0.014 & 35.4 \\
2 & 0.5 & 1 & 1  & 1.29 & 0.55 & 0.02 & 5000 & 0.020 & 25.0 \\
\hline
\end{tabular}
\caption{Initial conditions of the KHI problem setup for \texttt{Athena} simulations.  From {\it left} to {\it right} the columns are the Region number (see Figure \ref{fig:setup_sim}), shear flow velocity, gas pressure, gas density, sound speed, Mach number, strength of the magnetic field aligned with the shear flow, gas-to-magnetic pressures ratio, \Alfven speed, and \Alfven Mach number.  Where units apply but are left unspecified, these are arbitrary code units.}
\label{tab:initial_params}
\end{table}

\begin{table}
\renewcommand{\arraystretch}{1.33}
\centering
\begin{tabular}{l c c c c c c c}
\hline
\hline
ID$^{a}$ & $N$ & $t_{\rm start}^{d}$ & $L_{z}$ & $B_{0}$ & $\nu / \nu_{\rm fid}$ & $\eta / \eta_{\rm fid}$ \\
\hline
3M1024 & 1024 & $t_{0}$ & 1 & 0.02 & 0 & 0 \\
3M512 & 512 & $t_{0}$ & 1 & 0.02 & 0 & 0 \\
3M256 & 256 & $t_{0}$ & 1 & 0.02 & 0 & 0 \\
3M128 & 128 & $t_{0}$ & 1 & 0.02 & 0 & 0 \\
2M16384 & 16,384 & $t_{0}$ & 1 & 0.02 & 0 & 0 \\
2M8192 & 8192 & $t_{0}$ & 1 & 0.02 & 0 & 0 \\
2M4096 & 4096 & $t_{0}$ & 1 & 0.02 & 0 & 0 \\
2M2048 & 2048 & $t_{0}$ & 1 & 0.02 & 0 & 0 \\
2M1024 & 1024 & $t_{0}$ & 1 & 0.02 & 0 & 0 \\
2M512 & 512 & $t_{0}$ & 1 & 0.02 & 0 & 0 \\
2M256 & 256 & $t_{0}$ & 1 & 0.02 & 0 & 0 \\
2M128 & 128 & $t_{0}$ & 1 & 0.02 & 0 & 0 \\
3H512 & 512 & $t_{0}$ & 1 & 0 & 0 & 0 \\
2H8192 & 8192 & $t_{0}$ & 1 & 0 & 0 & 0 \\
3M512$^{2\nu}_{2\eta}$ & 512 & $t_{0}$ & 1 & 0.02 & 2 & 2 \\
3M512$^{2\nu}_{1\eta}$ & 512 & $t_{0}$ & 1 & 0.02 & 2 & 1 \\
3M512$^{1\nu}_{2\eta}$ & 512 & $t_{0}$ & 1 & 0.02 & 1 & 2 \\
3M512$^{1\nu}_{1\eta}$ & 512 & $t_{0}$ & 1 & 0.02 & 1 & 1 \\
3M512D$^{2\nu}_{2\eta}$ & 512 & $t_{\rm peak}$ & 1 & 0.02 & 2 & 2 \\
3M512D$^{1\nu}_{1\eta}$ & 512 & $t_{\rm peak}$ & 1 & 0.02 & 1 & 1 \\
3M256D$^{1\nu}_{1\eta}$ & 256 & $t_{\rm peak}$ & 1 & 0.02 & 1 & 1 \\
3M128D$^{1\nu}_{1\eta}$ & 128 & $t_{\rm peak}$ & 1 & 0.02 & 1 & 1 \\
3M512D$^{\nu/2}_{\eta/2}$ & 512 & $t_{\rm peak}$ & 1 & 0.02 & 1/2 & 1/2 \\
3M512D$^{\nu/4}_{\eta/4}$ & 512 & $t_{\rm peak}$ & 1 & 0.02 & 1/4 & 1/4 \\
3M512D$^{\nu/8}_{\eta/8}$ & 512 & $t_{\rm peak}$ & 1 & 0.02 & 1/8 & 1/8 \\
3M512J$^{b}$ & 512 & $t_{0}$ & 1 & 0.02 & 0 & 0 \\
3M512z2 & 512$^{c}$ & $t_{0}$ & 2 & 0.02 & 0 & 0 \\
3M512z4 & 512$^{c}$ & $t_{0}$ & 4 & 0.02 & 0 & 0 \\
\hline
\end{tabular}
\caption{Table of KHI simulations referred to in our study.  From {\it left} to {\it right} the columns are the simulation identification tag, grid resolution in each dimension, time at which the simulation was started from, size of the $z$-domain (in code units), initial magnetic field strength (in code units), kinematic viscosity coefficient relative to the fiducial value ($\nu_{\rm fid} = 2.6 \times 10^{-5}$), and Ohmic resistivity coefficient relative to the fiducial value ($\eta_{\rm fid} = 1.7 \times 10^{-5}$). \newline $^{a}$The ID tag generally follows a straightforward, three-part naming convention.  The first number indicates the dimensionality (i.e., 2D or 3D), the letter denotes the gas dynamics used (i.e., M for MHD or H for hydrodynamics), and the trailing number specifies the grid resolution in each direction. \newline $^{b}$The shear layer in this simulation was discontinuous, corresponding to the width parameter, $a = 0$.  All other simulations adopt $a = 0.01$ (in code units). \newline $^{c}$Simulations 3M512z2 and 3M512z4 have $N_{z} = 1024$ and $N_{z} = 2048$, respectively. \newline $^{d}$A start time of $t_{0}$ means the simulation started from the initial KHI configuration depicted in Figure \ref{fig:setup_sim}.  A start time of $t_{\rm peak}$ means the ideal MHD simulation evolved from $t_{0}$ to the point in the saturated state when the magnetic energy peaked and was then restarted with non-ideal MHD terms introduced.}
\label{tab:simlist}
\end{table}

\section{Spectral Analysis}
\label{sec:specanal}
\begin{figure}
  \includegraphics[width=84mm, trim=25mm 25mm 25mm 15mm, clip=true]{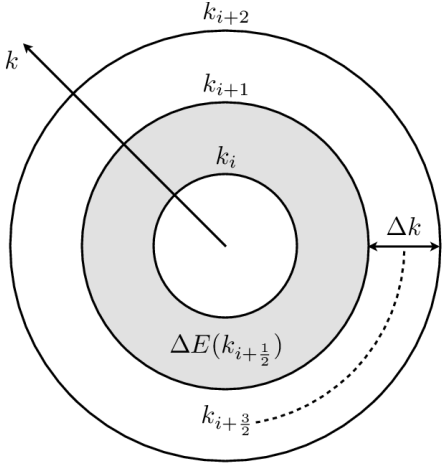}
  \caption{Schematic representation of how differential spectral energy density is spherically integrated over concentric $k$-shells of constant thickness, $\Delta k$.  For a given spectral energy density, $E(k)$, the differential spectral energy density, $\Delta E(k_{i+1/2})$, is computed as the sum-total energy contained within a shell of inner boundary $k_{i}$ and outer boundary $k_{i+1}$, where $\Delta k = k_{i+1} - k_{i}$ is constant.  To avoid double-counting, the spectral energy contained at the outer boundary of a shell (i.e., $E(k_{i+1})$ for $\Delta E(k_{i+1/2})$) is omitted, while the energy at the inner boundary is taken to be inclusive.  Spectral energy is integrated over shells of thickness $\Delta k L / (2 \pi) = 1$ from $k_{\rm min} L / (2 \pi)= 0$ to $k_{\rm max} L / (2 \pi) = k_{\rm Ny} L / (2 \pi) = N / 2$.  Plots of energy power spectra show the differential spectral energy density contained within a $k$-shell, $d E(k) / d k = \Delta E(k) / \Delta k$.}
  \label{fig:kshells}
\end{figure}

Throughout this work, we use energy power spectra to examine at which scales the majority of the magnetic energy is generated and how the spectral shape of the different energy reservoirs (i.e., kinetic, magnetic, and internal) evolve.  The kinetic, magnetic, and internal energy power spectra, also referred to as spectral energy densities, are defined as,
\begin{align}
E_{\rm K}(k) &= \frac{1}{2} \left( \widehat{\mathbf{v}}(k) \cdot \widehat{\left[ \rho \mathbf{v} \right]}^{\ast}(k) \right)\\
E_{\rm M}(k) &= \frac{1}{2} \left( \widehat{\mathbf{B}}(k) \cdot \widehat{\mathbf{B}}^{\ast}(k) \right) \\
E_{\rm I}(k) &= \frac{\widehat{P}(k)}{\gamma - 1},
\end{align}
where $k \equiv | \mathbf{k} | = \sqrt{k_{x}^{2} + k_{y}^{2} + k_{z}^{2}}$, an asterisk superscript denotes a complex conjugate, and $\widehat{F}(\mathbf{k})$ indicates the Fourier transform of the quantity $f(\mathbf{x})$,
\begin{equation}
\widehat{F}(\mathbf{k})= \iiint f(\mathbf{x}) e^{-i \mathbf{k} \cdot \mathbf{x}} d^{3} \mathbf{x}. \label{eq:Fouriertransform}
\end{equation}
No normalization is performed, as the magnitude of energy transfer is of interest, rather than merely the spectral shape. To improve statistics and aid in interpretation, the energy power spectra are integrated over concentric spherical shells of thickness, $\Delta k L / (2 \pi) = 1$, as shown schematically in Figure \ref{fig:kshells}.  This yields the differential power contained within a shell,
\begin{equation}
  \frac{d E(k)}{d k} = \frac{\Delta E(k)}{\Delta k},
\end{equation}
centered on half-integer values of wavenumber $k$.

Sometimes we will have cause to perform a shell-averaged normalization of spectral energy density (see \S \ref{sec:dissip}) according to,
\begin{equation}
\frac{d E(\langle k \rangle)}{d k} = \left[ \int k^{2} \frac{d E(k)}{dk} dk \right]^{-1} \frac{d E(k)}{d k},
\end{equation}
where the angled bracket convention, $d E \left( \langle k \rangle \right) / dk$, indicates that a shell average was performed on the spectral energy density.

Power spectra describe the distribution of energy across spatial scales; however, such distributions provide no clear way of determining how the energy transfers between scales and different forms (e.g., between magnetic and kinetic energies).  Transfer function analysis, first introduced by \citet{Kraichnan1967}, allows for the scale-by-scale quantification of energy transfer between reservoirs and identification of the mechanism responsible for the energy exchange.  The mechanics of deriving the transfer functions are given in Appendix \ref{append:transfunc}, with an outline of the derivations and an explanation of notation given here.  In this, we closely follow the approach and interoperation of \citet{PietarilaGraham2010}, who rigorously developed transfer analysis for compressible MHD in the context of the small-scale solar turbulent dynamo.  We specialize to the case of the KHI by decomposing the velocity field into contributions from the shearing flow and turbulent fluctuations \citep{FromangPapaloizou2007a, FromangPapaloizou2007b, SimonHawleyBeckwith2009},
\begin{equation}
\mathbf{v} = \mathbf{v}_{\rm sh} + \mathbf{v}_{\rm t} \label{eq:vdecomp},
\end{equation}
where $\mathbf{v}_{\rm t}$ is the turbulent velocity field and $\mathbf{v}_{\rm sh}$ is the background flow field, defined as,
\begin{equation}
\mathbf{v}_{\rm sh} = v_{\rm sh} \left( z \right) \widehat{\mathbf{y}} = \frac{\widehat{\mathbf{y}}}{L_{x} L_{y}} \iint v_{y}(x,y,z) dx dy.
\end{equation}

Inserting the decomposed velocity field into the momentum, energy, and induction equations, taking the Fourier transform, and performing the appropriate dot product (see Appendix \ref{append:transfunc}), the complete transfer function equations for kinetic, magnetic, and internal energies are,
\begin{align}
\frac{d E_{\rm K}(k)}{d t} =&~ T_{\rm IKC}(k) + S_{\rm IKC}(k) + T_{\rm KKA}(k) + X_{\rm KKA}(k) + \nonumber \\
&~ T_{\rm BKT}(k) + S_{\rm BKT}(k) + T_{\rm BKP}(k) + S_{\rm BKP}(k) + \nonumber \\
&~ T_{\rm KKC}(k) + S_{\rm KKC}(k) + X_{\rm KKC}(k) + D_{\rm K}(k) \label{eq:transfunc_KE} \\
\frac{d E_{\rm M}(k)}{d t} =&~ T_{\rm BBA}(k) + S_{\rm BBA}(k) + T_{\rm KBT}(k) + S_{\rm KBT}(k) + \nonumber \\
&~ T_{\rm KBP}(k) + D_{\rm M}(k) \label{eq:transfunc_ME} \\
\frac{d E_{\rm I}(k)}{d t} =&~ T_{\rm KIA}(k) + S_{\rm KIA}(k) + T_{\rm KIC}(k) + D_{\rm I}(k), \label{eq:transfunc_IE}
\end{align}
where the notation is described below.  These are time evolution equations of spectral energy densities.  Fourier transforms are computed according to Equation \ref{eq:Fouriertransform} using a fast Fourier transform algorithm.  The shear layer is not driven and is continually decaying; thus, the energy densities in the saturated state are not in a steady state and time derivatives are calculated explicitly.

Following  \citet{PietarilaGraham2010}, we interpret the transfer function $T,S,X_{\rm XYF}(k)$ as measuring the net energy transfer rate from {\it all scales} of reservoir X to scale $k$ of reservoir Y, where the energy exchange is via the force F.  The net energy transfer from reservoir X into reservoir Y at scale $k$ is positive (negative) for $T,S,X_{\rm XYF}(k) > 0~(< 0)$. The available energy reservoirs are kinetic (K), magnetic (M), and internal (I). The mediating forces (F) depend on the exact form of each transfer function, but in general these forces are compressive motions (C), advection (A), magnetic tension (T), and magnetic pressure (P). Energy transfer due to purely turbulent motions, $\mathbf{v}_{\rm t}$, is denoted by $T_{\rm XYF}(k)$; the background shear flow, $\mathbf{v}_{\rm sh}$, by $S_{\rm XYF}(k)$; or some hybrid cross term involving both $\mathbf{v}_{\rm t}$ and $\mathbf{v}_{\rm sh}$ by $X_{\rm XYF}(k)$. Finally, the terms $D_{\rm K}(k)$, $D_{\rm M}(k)$, and $D_{\rm I}(k)$ in Equations \ref{eq:transfunc_KE}--\ref{eq:transfunc_IE} are simply the residuals of the time derivative of spectral energy density and the sum of all transfer functions, resulting in a measure of numerical dissipation rate as a function of scale \citep{FromangPapaloizou2007a, FromangPapaloizou2007b, SimonHawleyBeckwith2009}.

All transfer functions are spherically integrated over shells of constant thickness $\Delta k L / (2 \pi) = 1$ and then plotted as $k \cdot (d T_{\rm XYF}(k)/ d k)$ {\it versus} ${\rm log}(k)$ so that the peak in the spectrum corresponds to the wavenumber containing the most power \citep{ZdziarskiGierlinski2004}.  We choose to time-average the transfer functions over the same intervals shown in the energy power spectral analysis of Figure \ref{fig:powspec_allE}.  This improves statistics across all $k$ and allows us to make robust statements regarding energy exchange during different stages of the KHI evolution.

\section{Convergence}
\label{sec:converge}
\begin{figure}
  \centering
  \includegraphics[width=\columnwidth]{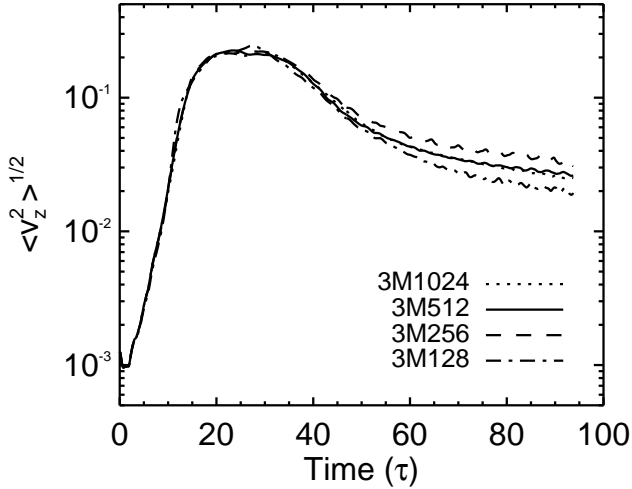}
  \caption{Volume-averaged rms velocity component transverse to the shear layers for the resolved shearing runs 3M1024 ({\it dotted line}), 3M512 ({\it solid line}), 3M256 ({\it dashed line}), and 3M128 ({\it dash-dot line}).  Convergence is demonstrated in the linear growth regime for all resolutions considered here, but $\langle v_{z}^{2} \rangle^{1/2}$ cannot be used as a diagnostic of convergence in the decaying regime.  Here, and in all subsequent figures, time is parameterized in units of the linear growth e-folding time, $\tau$, as computed from the fiducial 3M512 simulation.}
  \label{fig:Vz_rms}
\end{figure}

In the absence of explicit dissipative terms in the conservation equations \ref{eq:continuity}--\ref{eq:induction}, the effective (i.e., numerical) dissipation present in the simulation is governed by the choice of grid resolution.   The numerical dissipation, expressed in units of diffusivity as $(\Delta \mathbf{x})^{2} / \Delta t$, decreases with improved grid resolution for a fixed timestep.  As grid resolution elements become finer, small-scale structures are preserved that would otherwise be smeared out by under-resolved simulations whose numerical dissipation scale is too large to capture said structures.  Small-scale structure can drive energy exchange and morphological evolution, particularly in the non-linear and turbulent regimes.  Therefore, establishing a converged solution is paramount for a physical interpretation of the simulation results.

Convergence, in the formal sense, refers to an unchanging power spectrum across all scales when resolution is increased.  However, this is unattainable in inviscid turbulent simulations.  Expecting an exact point-to-point match of a quantity between different grid resolutions is inappropriate in the non-linear regime given the turbulent nature of the problem at hand and the presence of numerical dissipation.  Instead, we refer to convergence in the sense that quantities integrated over the entire volume do not change appreciably for a factor of two increase in $N$, the grid resolution in each dimension.  This definition we adopt is colloquially referred to as {\it virtual convergence} and is demonstrated as an effective diagnostic in practice \citep[e.g., ][]{Palottietal2008, LemasterStone2009}.

Convergence of the linear growth stage of the KHI can be assessed through the volume-averaged root mean square (rms) velocity transverse to the shear layer, $\langle v_{z}^{2} \rangle^{1/2}$ \citep[e.g.,][]{Franketal1996}.  In this work, angled brackets surrounding a quantity denote volume averages, where the volume-average of quantity $Q(x,y,z)$ is given by,
\begin{equation}
\left< Q \right> = \frac{1}{L_{x} L_{y} L_{z}} \iiint Q \left( x,y,z \right) dx dy dz. \label{eq:Vavg}
\end{equation}
When considering the volume-average of a vector quantity, such as the magnetic field, $\mathbf{B}$, we consider the magnitude of that vector quantity, $B = \left| \mathbf{B} \right|= \sqrt{B_{x}^{2} + B_{y}^{2} + B_{z}^{2}}$, and take its volume-average.

Figure \ref{fig:Vz_rms} shows $\langle v_{z}^{2} \rangle^{1/2}$ for the runs 3M128, 3M256, 3M512, and 3M1024, where the initial value is dictated by the perturbation of the equilibrium configuration.  The linear growth phase of the KHI corresponds to the exponentially increasing portion of $\langle v_{z}^{2} \rangle^{1/2}$.  We choose to parameterize time in terms of the linear growth e-folding time, $\tau \simeq 0.16 t$.  We evaluate $\tau$ over the exponentially increasing portion of $\langle v_{z}^{2} \rangle^{1/2}$ according to, $\langle v_{z}^{2} \rangle^{1/2} = A e^{t / \tau}$, where $A$ is the initial rms transverse velocity and $t$ is time in code units.  Although $\tau$ is decreasingly relevant as the flow becomes turbulent, it is physically motivated and well-defined during the linear growth.  Figure \ref{fig:Vz_rms} shows that the linear growth phase converges even at the lowest 3D resolution considered here, $N = 128$. The linear growth phase of the instability terminates at $\tau \simeq 20$, following which $\langle v_{z}^{2} \rangle^{1/2}$ saturates and then decays for $\tau \gtrsim 30$. During this phase of the evolution, $\langle v_{z}^{2} \rangle^{1/2}$ exhibits non-monotonic behaviour with resolution. As a result, we conclude that $\langle v_{z}^{2} \rangle^{1/2}$ is not a sufficient diagnostic to determine convergence in the non-linear regime.  From here onward, we take convergence to mean in the virtual sense described above.

\begin{figure}
  \centering
  \includegraphics[width=\columnwidth]{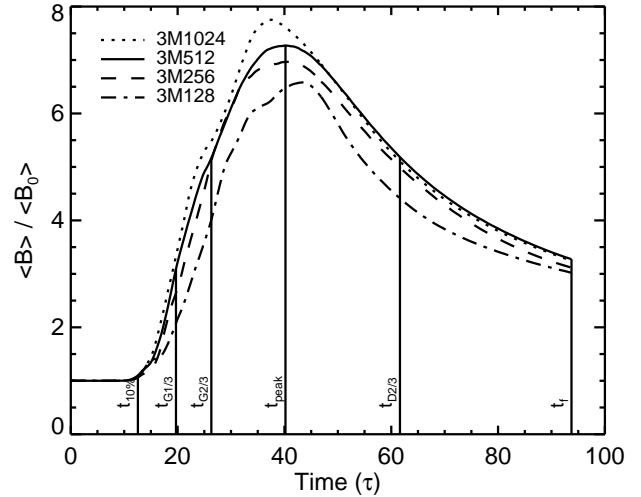}
  \caption{Time evolution of 3D MHD simulations showing the amplification, saturation, and decay of the volume-averaged magnetic field magnitude, $\left<B\right>$, relative to the initial volume-averaged magnetic field magnitude, $\left<B_{0}\right>$.  The 3D simulations shown in this convergence study are 3M128 ({\it dash-dot line}), 3M256 ({\it dashed line}), 3M512 ({\it solid line}), and 3M1024 ({\it dotted line}).  The times marked by {\it vertical lines} correspond to the 3M512 simulation and are defined in the text (see \S \ref{sec:devevo}).  Convergence is demonstrated at a resolution, $N = 512$, for the linear growth and non-linear decay regimes, with only modest differences in the saturation amplitude.}
  \label{fig:Bamp}
\end{figure}

\begin{figure}
  \centering
  \includegraphics[width=\columnwidth]{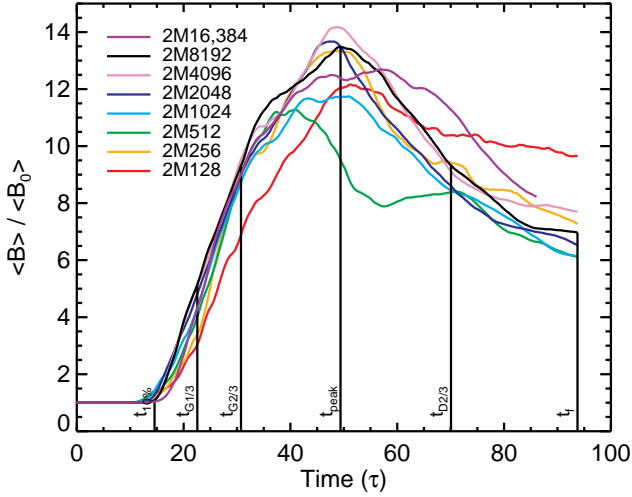}
  \caption{Time evolution of 2D MHD simulations showing the amplification, saturation, and decay of the volume-averaged magnetic field magnitude, $\left<B\right>$, relative to the initial volume-averaged magnetic field magnitude, $\left<B_{0}\right>$.  The 2D simulations shown in this convergence study are 2M128 ({\it red line}), 2M256 ({\it orange line}), 2M512 ({\it green line}), 2M1024 ({\it cyan line}), 2M2048 ({\it blue line}), 2M4096 ({\it violet line}), 2M8192 ({\it black line}), and 2M16384 ({\it purple line}).  The times marked by {\it vertical lines} correspond to the 2M8192 simulation and are defined in the text (see \S \ref{sec:devevo}).  While the linear regime is well-converged at a resolution, $N = 512$, neither the non-linear saturated state nor the late-time decay show indications of convergence, even for the very high resolution case, $N = 16,384$.}
  \label{fig:Bamp2D}
\end{figure}

\begin{figure}
  \centering
  \includegraphics[width=\columnwidth]{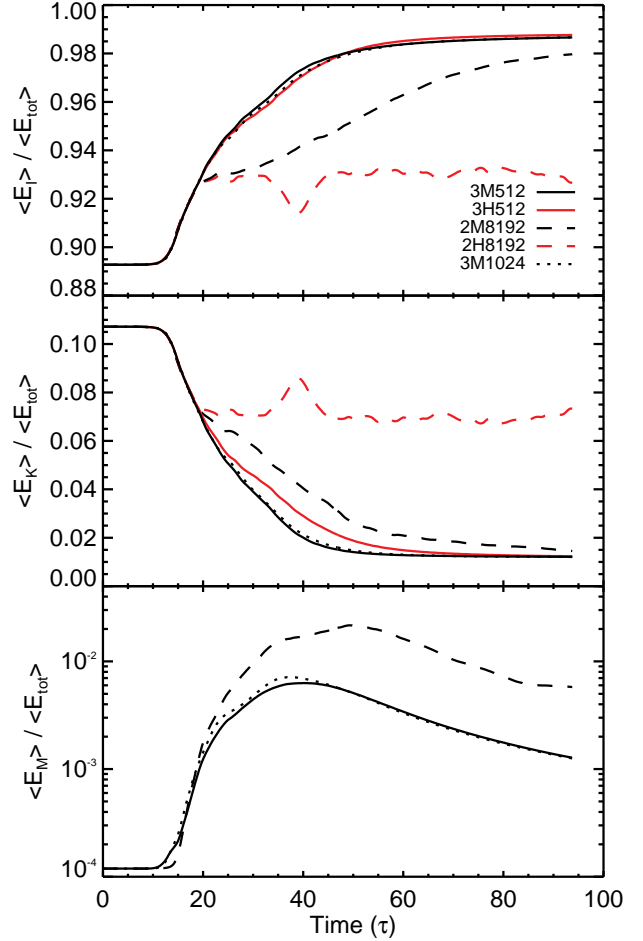}
  \caption{Time evolutions of the volume-averaged quantities: internal energy, $\left<E_{\rm I}\right>$ ({\it top panel}), kinetic energy, $\left<E_{\rm K}\right>$ ({\it middle panel}), and magnetic energy, $\left<E_{\rm M}\right>$ ({\it bottom panel}).  All volume-averaged energies are shown relative to the total volume-averaged energy in the computational box, $\left<E_{\rm tot}\right>$.  Shown are results for KHI simulations 3M1024 ({\it black dotted lines}), 3M512 ({\it black solid lines}), 3H512 ({\it red solid lines}), 2M8192 ({\it black dashed lines}), and 2H8192 ({\it red dashed lines}).  Magnetic energy is more efficiently generated from the available kinetic energy and less efficiently dissipated for the 2D KHI.}
  \label{fig:EVavgTevol}
\end{figure}

Figure \ref{fig:Bamp} shows the time evolution of the volume-averaged magnitude of the magnetic field for sets of 3D MHD simulations at various grid resolutions.  This figure serves as a convergence study of the KHI simulations in the non-linear regime, with convergence obtained at $N = 512$. That is, the difference in evolution of magnetic energy in the non-linear regime of interest (i.e., $\tau \gtrsim 50$) between the $N = 512$ and $N = 1024$ simulations is at the 1\% level. Based on Figure \ref{fig:Bamp} and the discussion presented above, we conclude that the decay of turbulence in the non-linear regime is driven by physical, rather than numerical processes. This will be confirmed by the transfer function analysis of \S \ref{sec:transfunc} and \S \ref{sec:dissip}. We therefore treat $N = 512$ as our fiducial resolution and the 3M512 run as our fiducial simulation.

By contrast, 2D MHD simulations of the KHI do not exhibit convergence in the non-linear regime.  Figure \ref{fig:Bamp2D} shows the same quantity as in Figure \ref{fig:Bamp}, but for a series of 2D simulations of the KHI at resolutions up to $N = 16,384$, with little indication of convergence in the non-linear regime. This is evidenced by the absence of both a consistent peak in the magnetic field and a single sustained value at late times.  A more detailed comparison of the evolution of the two- and three-dimensional KHI is found in Figure \ref{fig:EVavgTevol}, which shows the time evolution of the volume-averaged internal, kinetic, and magnetic energies, each normalized by the volume-averaged total energy. Inspecting Figure \ref{fig:EVavgTevol}, the evolution of the energetics in the 2D system shows substantial differences from the 3D case. In particular, the 2D flow is more efficient than the 3D flow at generating magnetic energy from the available kinetic energy and this magnetic energy is less efficiently dissipated into heat in the 2D case.  Internal energy is the dominant energy component and increases throughout the simulation because there is no cooling prescription.

\begin{figure}
  \includegraphics[width=\columnwidth]{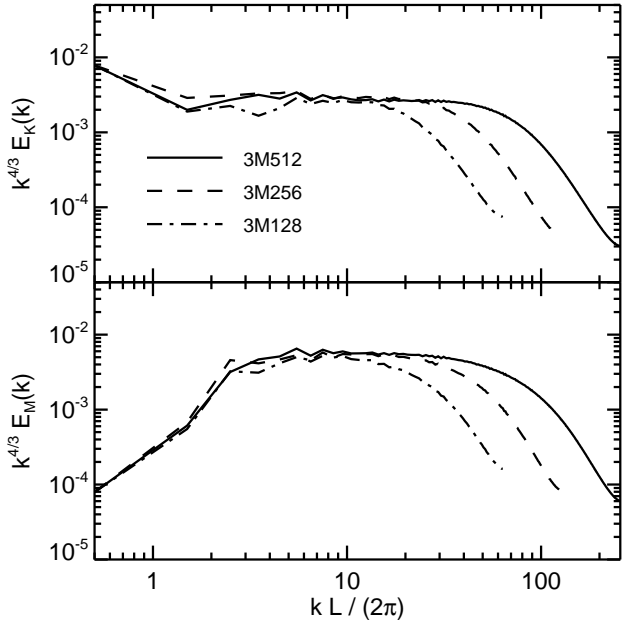}
  \caption{Time-averaged, one-dimensional spectral energy densities for simulations 3M128 ({\it dash-dot lines}), 3M256 ({\it dashed lines}), and 3M512 ({\it solid lines}).  Time averages are performed over the non-linear turbulent decay interval, [$t_{\rm peak}$, $t_{\rm f}$].  {\it Top panel}:  Kinetic energy power spectra, $E_{\rm K}(k)$.  {\it Bottom panel}:  Magnetic energy power spectra, $E_{\rm M}(k)$.  Spectral energy densities are compensated by a factor of $k^{4/3}$.  Consistent behaviour in both $E_{K}(k)$ and $E_{\rm M}(k)$ is seen in the non-linear turbulent decay regime across the resolutions studied, with an inertial range emerging at intermediate scales for the 3M512 simulation.  The effect of increasing resolution is to shift the dissipation scale to smaller scales (i.e., higher $k$).}
  \label{fig:powspec_3Dres}
\end{figure}

\begin{figure}
  \includegraphics[width=\columnwidth]{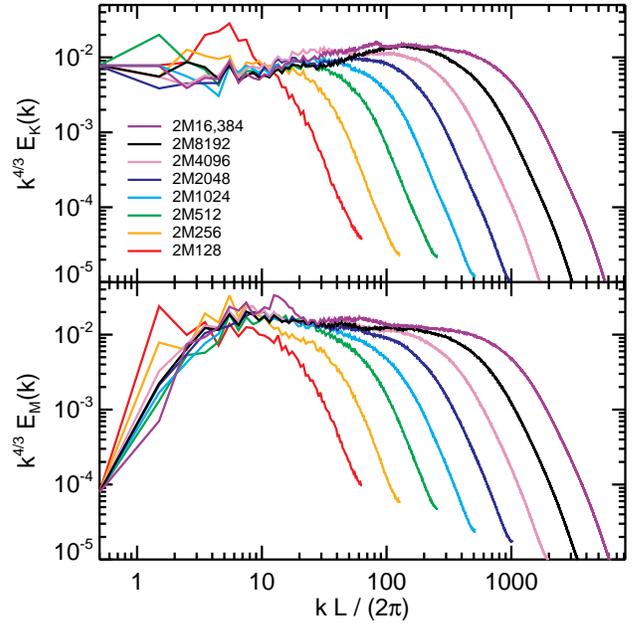}
  \caption{Time-averaged, one-dimensional spectral energy densities for numerous 2D MHD simulations at different resolutions.  Time averages are performed over the non-linear turbulent decay interval, [$t_{\rm peak}$, $t_{\rm f}$].  Line colours are the same as in Figure \ref{fig:Bamp2D} and the resolution hierarchy can be deduced from the cut-off in $k$.  {\it Top panel}: Kinetic energy power spectra, $E_{\rm K}(k)$.  {\it Bottom panel}: Magnetic energy power spectra, $E_{\rm M}(k)$.  Spectral energy densities are compensated by a factor of $k^{4/3}$.  With increased resolution, an approximately converged inertial range emerges in $E_{\rm M}(k)$ and the dissipation scale moves to higher $k$.  However, increasing resolution alters the spectral distribution of kinetic energy, rather than merely pushing the dissipation scale to smaller scales.}
  \label{fig:powspec_2Dres}
\end{figure}

Further evidence of the differences in behaviour of the two- and three-dimensional KHI can be found through comparison of Figures \ref{fig:powspec_3Dres} and \ref{fig:powspec_2Dres}. These figures show the time-averaged spectral distributions of the magnetic and kinetic energies in the three- and two-dimensional simulations, respectively, compensated by $k^{4/3}$ to enable visual comparison.  Figure \ref{fig:powspec_3Dres} shows that, for the 3D simulations, the spectral distribution of these quantities follow a $k^{-4/3}$ power-law for $k L / (2 \pi) \gtrsim 3$, over all the resolutions considered.  The main effect of increasing resolution is to move the dissipation scale to progressively smaller scales; from $k L / (2 \pi) \sim 10$ ($N = 128$) to $k L / (2 \pi) \sim 50$ ($N = 512)$. As evidenced by Figure \ref{fig:powspec_2Dres}, the behaviour of the 2D simulations is different. While the spectral distribution of the magnetic energy reservoir shows approximate convergence to a $k^{-4/3}$ power-law, the spectral distribution of the kinetic energy does not appear to follow a simple power-law and shows a changing dependence as higher resolutions are probed. That is, the effect of increasing resolution in the 2D case is to alter the spectral distribution of kinetic energy, rather than to simply move the dissipation scale to higher $k$ as observed in the 3D case. This suggests the operation of an inverse cascade\footnote{By ``inverse cascade'' we mean that energy in the magnetic reservoir is initially spectrally dominated at small scales and evolves to become primarily distributed on large scales.  We do not mean to imply a dynamo process by using this phrase.} in decaying 2D magnetized turbulence, which is already known to occur in 2D hydrodynamic turbulence.  In nature, turbulence must be inherently 3D; therefore, we restrict the remainder of the analysis to the 3D KHI simulations.

\section{Evolution}
\label{sec:devevo}
Here, we explore in detail the evolution of the non-driven KHI with a focus on the properties of the non-linear MHD turbulent regime.  We start with simple volume-averaged quantities to characterize global properties and morphology (\S \ref{sec:morph}) and then use spectral energy densities to quantify the distribution of energy across spatial scales (\S \ref{sec:powspec}).  Increasing the utility of our analysis diagnostic, we take advantage of spectral energy transfer function analysis to probe deep into the physics of decaying MHD turbulence (\S \ref{sec:transfunc}) and later we apply this tool to study dissipation (\S \ref{sec:dissip}).  The transfer function diagnostic allows for an in-depth quantification of MHD turbulence in general and here we demonstrate its power by focusing on a well-studied problem --- the KHI.

\subsection{Global Properties and Morphology}
\label{sec:morph}
\begin{figure*}
  \centering
  \includegraphics[width=2.0\columnwidth]{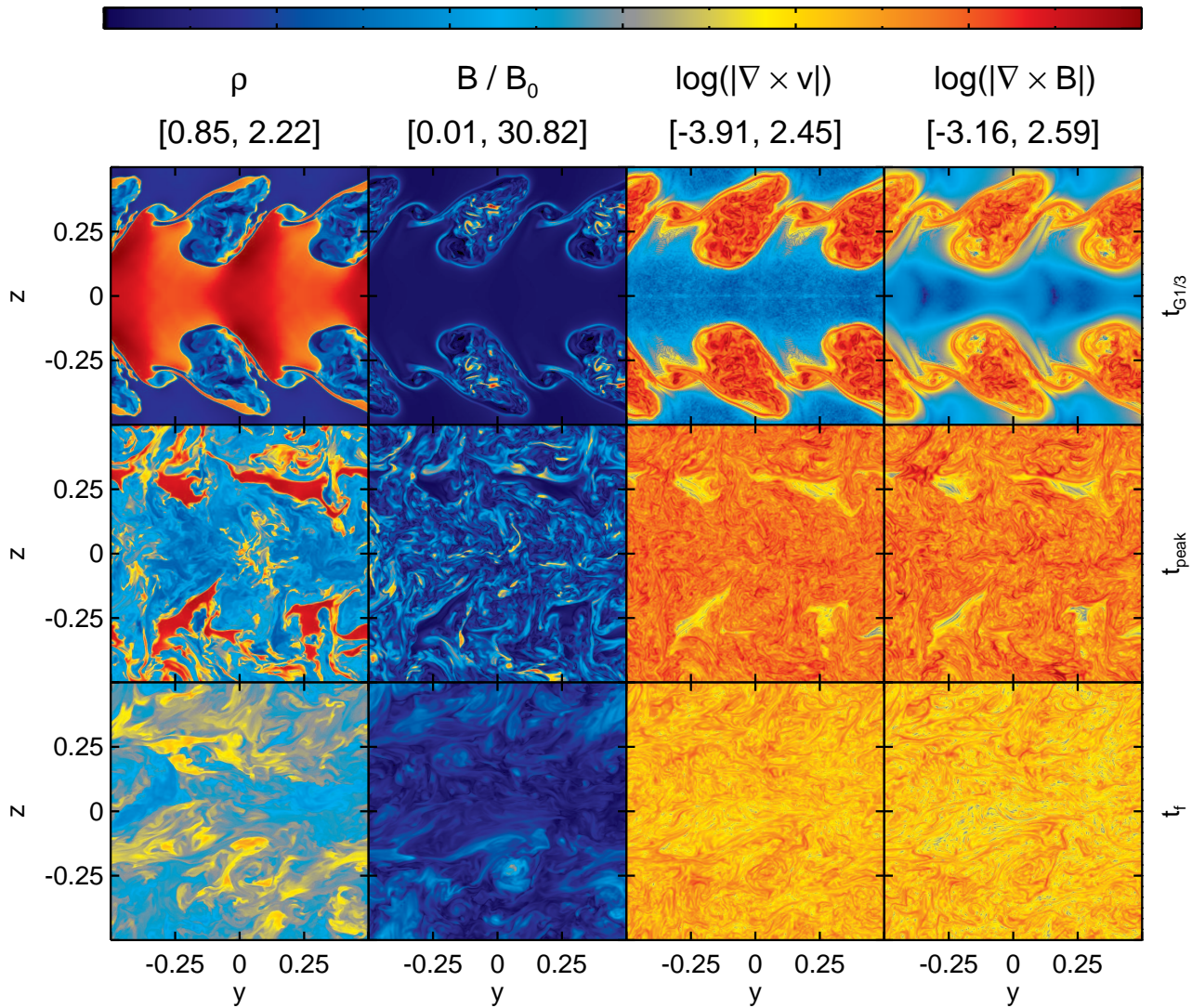}
  \caption{2D slices taken in the $yz$-plane at $x = 0.5 L$ from the 3M512 simulation at times $t_{\rm G1/3}$ ({\it first row}), $t_{\rm peak}$ ({\it second row}), and $t_{\rm f}$ ({\it third row}).  From {\it left} to {\it right}, the columns show the gas density, $\rho$, magnetic field strength relative to the initial value, $B / B_{0}$, logarithm of the vorticity magnitude, ${\rm log}(\left| \nabla \times \mathbf{v} \right|)$, and logarithm of the current density magnitude, ${\rm log}(\left| \nabla \times \mathbf{B} \right|)$.  The bracketed numbers above each column mark the [minimum, maximum] parameter values for the linear-scale color bar used to plot the respective quantity in all rows of the column.  By the saturation time, $t_{\rm peak}$, the initial shear layer is destroyed and the flow enters a state of decaying turbulence.}
  \label{fig:yz_slice}
\end{figure*}

Comparing Figures \ref{fig:Vz_rms} and \ref{fig:Bamp} shows that significant magnetic field amplification only occurs within the non-linear stage of the instability (i.e., for $\tau \gtrsim 10$).  Figure \ref{fig:Bamp} highlights that there are three regimes in the evolution of the magnetic field during the non-linear stage: amplification ($10 \lesssim \tau \lesssim 30$), saturation ($30 \lesssim \tau \lesssim 50$) and decay ($\tau \gtrsim 50$). For purposes of clarity during subsequent discussion, we further subdivide these regimes into intervals marked by {\it vertical lines} in Figure \ref{fig:Bamp}. These lines represent, in chronological order, the times during the fiducial simulation, 3M512, at which the volume-averaged magnetic field magnitude relative to the initial field strength grows by 10\% ($t_{10\%} = 2.01 $), grows to $\frac{1}{3} B_{\rm peak}$ ($t_{\rm G1/3} = 3.15$), grows to $\frac{2}{3} B_{\rm peak}$ ($t_{\rm G2/3} = 4.21$), reaches $B_{\rm peak}$ ($t_{\rm peak} = 6.43$), decays to $\frac{2}{3} B_{\rm peak}$ ($t_{\rm D2/3} = 9.86$), and the time at the termination of the simulation ($t_{\rm f} = 15.00$). The maximum amplitude attained by the magnetic field in the fiducial simulation is $B_{\rm peak} = 7.27 B_{0}$ and the subscripts on the times are meant to indicate growth (G) and decay (D) stages of the magnetic energy. 

To visually assess the development of the KHI, slices of gas density, magnetic field magnitude, vorticity magnitude, and current density magnitude are shown in Figure \ref{fig:yz_slice} for the fiducial simulation.  At $t = t_{10\%}$ (not shown in Figure \ref{fig:yz_slice}), the familiar linear growth wave-like pattern of the KHI is developed, with numerous small-scale, low-pressure vortices forming that are the sites of magnetic field amplification due to twisting and stretching of field lines.  Although initially less pronounced than the small-scale vortices, a set of two large vortices along each shear layer begins to develop as a consequence of the single mode perturbation introduced into the computational box at $t_{0} = 0$.  The KHI continues to develop into the non-linear stage by $t = t_{\rm G1/3}$, at which time two primary commensurate features have been established along each interface with multiple mini-vortices arising from secondary instabilities.  The non-linear evolution continues and produces finger-like strands of density, magnetic field, and pressure by $t = t_{\rm G2/3}$ (not shown in Figure \ref{fig:yz_slice}).  When the magnetic field reaches its peak amplitude at $t = t_{\rm peak}$, the shear layer is nearly shredded beyond recognition into turbulence with evidence for the single mode form of the initial perturbation also nearly erased.  At this point in time, the magnetic energy production mechanism (i.e., a driven shear layer) is absent and the magnetic field begins to decay as the fluid motions remain turbulent and the fluid is well-mixed.  Subsequent evolution of the system to late times is characterized by gradual decay of magnetic energy.

\subsection{Spectral Energy Densities}
\label{sec:powspec}
\begin{figure}
  \centering
  \includegraphics[width=\columnwidth]{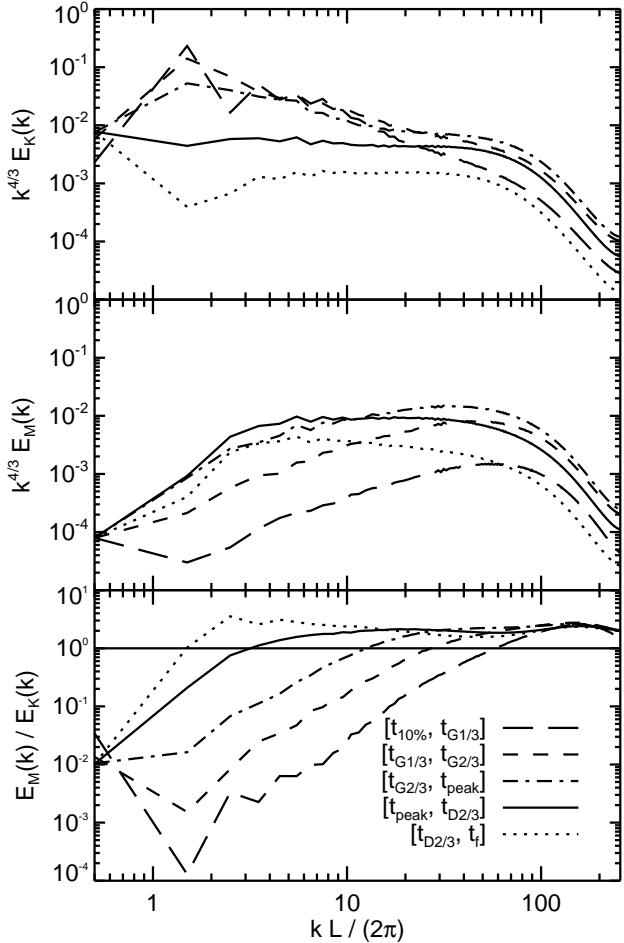}
  \caption{Time-averaged, one-dimensional spectral energy densities for the 3M512 simulation.  From {\it top} to {\it bottom} are the power spectra for kinetic energy, magnetic energy, and ratio of magnetic-to-kinetic energies.  Time averages are performed over the intervals $[t_{10\%},t_{\rm G1/3}]$ ({\it long-dashed lines}), $[t_{\rm G1/3},t_{\rm G2/3}]$ ({\it short-dashed lines}), $[t_{\rm G2/3},t_{\rm peak}]$ ({\it dash-dot lines}), $[t_{\rm peak},t_{\rm D2/3}]$ ({\it solid lines}), and $[t_{\rm D2/3},t_{\rm f}]$ ({\it dotted lines}).  As the simulation progresses, the spectral equipartition point shifts to larger scales, where magnetic energy dominates kinetic energy on scales smaller than the equipartition scale.}
    \label{fig:powspec_allE}
\end{figure}

The time-averaged kinetic and magnetic energy power spectra computed over these same time intervals from the fiducial simulation are shown in Figure \ref{fig:powspec_allE}.  Table \ref{tab:specslopes} provides the spectral slopes for the intermediate scales, $5 \le k L / (2 \pi) \le 30$, corresponding to each time averaging interval in Figure \ref{fig:powspec_allE}.  Figure \ref{fig:powspec_allE} shows that magnetic energy is concentrated in small spatial scales as the KHI begins to develop from the linear to saturated state (i.e., from $t = t_{10\%}$ to $t = t_{\rm peak}$).  As the volume-averaged magnetic field is amplified and peaks, magnetic energy at small scales (i.e., large $k$) grows with a fixed slope, while the spectral shape at small $k$ flattens. By contrast, during this phase of the non-linear evolution, kinetic energy contained on large scales, $k L / (2 \pi) \lesssim 30$, retains the same spectral slope and magnitude, while the majority of the kinetic energy amplification occurs on small spatial scales, $k L / (2 \pi) \gtrsim 30$, due to the development of small-scale vortices. Once the peak magnetic energy is reached, magnetic energy on large scales decays more gradually than that on small scales, causing the magnetic energy spectrum to steepen.  Conversely, kinetic energy on large scales decays more rapidly than that on small scales, causing the kinetic energy spectrum to flatten.

\begin{table}
\addtolength{\tabcolsep}{-1.5pt}
\centering
\begin{tabular}{c c c c c c}
\hline
\hline
$E(k)$ & $m^{10\%}_{\rm G1/3}$ & $m^{\rm G1/3}_{\rm G2/3}$ & $m^{\rm G2/3}_{\rm peak}$ & $m^{\rm peak}_{\rm D2/3}$ & $m^{\rm D2/3}_{\rm f}$ \\
\hline
$E_{\rm K}(k)$ & $-2.79(4)$ & $-2.27(3)$ & $-1.97(5)$ & $-1.47(2)$ & $-1.33(2)$ \\
$E_{\rm M}(k)$ & $-0.31(2)$ & $-0.44(3)$ & $-0.80(2)$ & $-1.33(2)$ & $-1.62(2)$ \\
\hline
\end{tabular}
\caption{Slopes, $m$, of the kinetic, $E_{\rm K}(k)$, and magnetic, $E_{\rm M}(k)$, one-dimensional spectral energy densities from a log-log fit over the range, $5 \le k L / (2 \pi) \le 30$, for the fiducial 3M512 simulation.  Time averaging intervals for the spectral energy densities are denoted by the subscript and superscript on $m$ and conform to the notation described in the text (see \S \ref{sec:devevo}).  Uncertainties on the last significant digit are given in parentheses and correspond to the $1\sigma$ level.  At late times, both $E_{\rm K}(k)$ and $E_{\rm M}(k)$ exhibit an inertial range approximated by a $k^{-4/3}$ power-law.}
\label{tab:specslopes}
\end{table}

The {\it bottom panel} of Figure \ref{fig:powspec_allE} makes the comparison of the spectral evolution of magnetic and kinetic energies explicit.  As the KHI develops from the linear regime toward the turbulent regime, the $E_{\rm M}(k) / E_{\rm K}(k)$ spectrum tends to increase and level off with increasing $k$.  This behaviour was also observed in the relativistic MHD KHI simulations of \citet{ZhangMacFadyenWang2009}.  The equipartition point of magnetic and kinetic energies slides toward larger scales for the entirety of the evolution, until magnetic energy dominates over kinetic energy across nearly all scales at the termination of the simulation.  Although the individual kinetic and magnetic energy spectra are decreasing in amplitude after $t = t_{\rm peak}$, the equipartition point continues to shift toward lower $k$.

\subsection{Spectral Energy Transfer Function Analysis}
\label{sec:transfunc}
\begin{figure}
  \centering
  \includegraphics[width=\columnwidth]{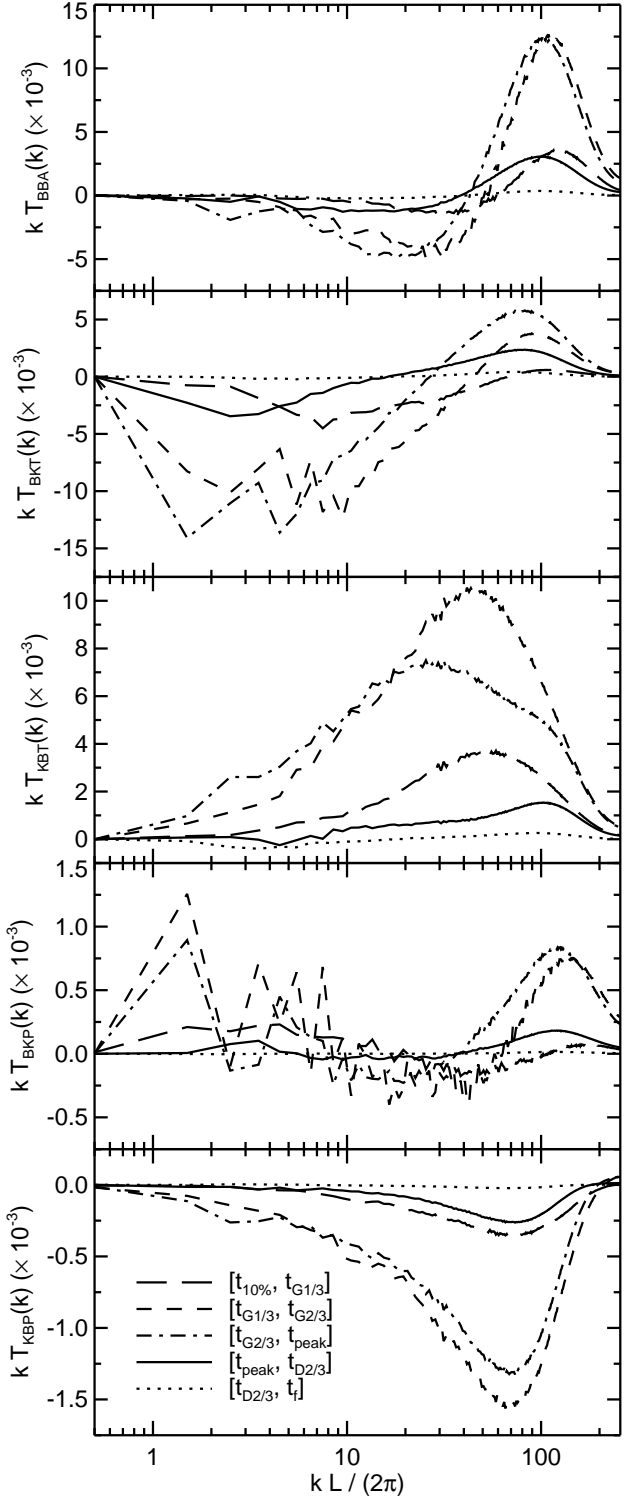}
  \caption{Time-averaged, one-dimensional spectral energy transfer functions associated with energy transfer into/out of the magnetic energy reservoir for the 3M512 simulation.  From {\it top} to {\it bottom} are the transfer functions $T_{\rm BBA}(k)$, $T_{\rm BKT}(k)$, $T_{\rm KBT}(k)$, $T_{\rm BKP}(k)$, and $T_{\rm KBP}(k)$.  Time averages are performed over the same intervals as in Figure \ref{fig:powspec_allE} with the same line style convention used.  The exact details of the energetics are seen to be highly time-dependent.  In general, the transfer function amplitudes evolve in time, but their spectral shape remains fairly consistent, albeit with some translation in $k$.}
  \label{fig:Txbx}
\end{figure}

The time-averaged transfer functions associated with energy exchange with the magnetic energy reservoir are shown in Figure \ref{fig:Txbx} and provide a quantification of magnetic energy sources/sinks as a function of scale $k$.  Only the transfer functions associated with turbulent motions (i.e., the $\mathbf{v}_{\rm t}$ piece of the velocity decomposition) are plotted, as we found that the transfer functions associated with pure shear (i.e., $S_{\rm XYF}$) and cross terms (i.e., $X_{\rm XYF}$) are negligible players in energy transfer in comparison.

We first consider energy transfer during the stages of the KHI development leading up to saturation (i.e., from $t = t_{10\%}$ to $t = t_{\rm peak}$).  The dominant growth mechanism of magnetic energy at large and intermediate scales is due to turbulent motions twisting/stretching magnetic field lines (i.e.,  $T_{\rm BKT}(k) < 0$ and $T_{\rm KBT}(k) > 0)$.  Transfer inside the magnetic energy reservoir by turbulent velocities (i.e., $T_{\rm BBA}(k)$) is responsible for an inverse cascade of magnetic energy.  Work done against magnetic pressure gradients by turbulent compressive motions (i.e., $T_{\rm BKP}(k)$ and $T_{\rm KBP}(k)$) is negligible in comparison to other magnetic transfer mechanisms.  Although not plotted, we inspected the kinetic energy transfer functions and found the following behaviour.  The dominant contribution to large-scale kinetic energy growth between $t = t_{10\%}$ and $t = t_{\rm peak}$ is from advection\footnote{Here, and henceforth, ``advection" is used to refer to the transfer of energy between scales but within the same form.  For example, the ``advection" of kinetic energy from large-to-small scales.} within the kinetic energy reservoir (i.e., $T_{\rm KKA}(k)$ and $X_{\rm KKA}(k)$).  Ancillary contributions come from both compressible turbulent motions within the kinetic energy reservoir (i.e., $T_{\rm KKC}(k)$ and $X_{\rm KKC}(k)$) and transfer from the internal energy reservoir by compression (i.e., $T_{\rm IKC}(k)$ and $S_{\rm IKC}(k)$).  Meanwhile, energy is being transferred out of the kinetic energy reservoir on these large scales into the magnetic energy reservoir by turbulent fluid motions twisting/stretching the magnetic field (i.e., $T_{\rm BKT}(k) < 0$).  The small-scale growth of kinetic energy during this time is overwhelmingly dominated by the same magnetic tension force\footnote{Here, and henceforth, ``tension'' is used to describe the restoring force directed along the radius of curvature that is exerted by bent magnetic field lines.  We do not mean to imply that the magnetic field is always putting tension on the fluid.} acting on turbulent motions that causes kinetic energy loss on large scales.  To summarize the KHI evolution leading up to saturation, we find that the magnetic field grows first at small scales and then cascades to larger scales, which is evidence for an inverse cascade operating in the KHI.  The dominant energy exchange mechanism involves turbulent fluid motions interacting with magnetic tension.

\begin{figure}
  \centering
  \includegraphics[width=84mm]{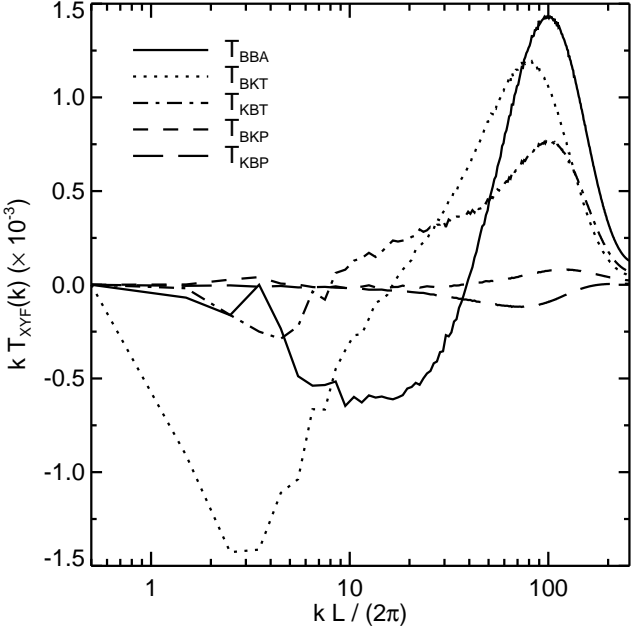}
  \caption{Time-averaged, one-dimensional spectral energy transfer functions associated with energy transfer into/out of the magnetic energy reservoir for the 3M512 simulation for the time-averaging interval $[t_{\rm peak}, t_{\rm f}]$.  The fluid is in a state of decaying turbulence over this interval in time.  The lines shown here are a representation of the data from the {\it solid lines} and {\it dotted lines} in Figure \ref{fig:Txbx}, but placed on the same scale to allow for relative comparisons.  The transfer functions shown are $T_{\rm BBA}(k)$ ({\it solid line}), $T_{\rm BKT}(k)$ ({\it dotted line}), $T_{\rm KBT}(k)$, ({\it dash-dot line}), $T_{\rm BKP}(k)$, ({\it short-dashed line}), and $T_{\rm KBP}(k)$, ({\it long-dashed line}).  Energy transfer is dominated by exchanges within the magnetic energy reservoir (i.e., $T_{\rm BBA}(k)$) and transfer mediated by magnetic tension (i.e., $T_{\rm BKT}(k)$ and $T_{\rm KBT}(k)$).  Magnetic pressure effects are small in comparison due to the weakly compressive nature of the subsonic and sub-Alf\'enic flow studied here.}
  \label{fig:Txbx_last}
\end{figure}

We now turn our attention to times after $t = t_{\rm peak}$, where the fluid is in a turbulent state and energy decays away by numerical dissipation.  From $t = t_{\rm peak}$ onward, when the simulation box is fully turbulent and the shear layer is destroyed, a transition occurs where the subsequent evolution of the kinetic energy spectrum over all scales is determined primarily by interactions with the magnetic energy reservoir.  Figure \ref{fig:Txbx_last} shows the energy transfer between magnetic energy transfer functions in the time-averaged decay stage from $t = t_{\rm peak}$ to $t = t_{\rm f}$.  As before, exchanges within the magnetic energy reservoir and transfer mediated by magnetic tension dominate the magnetic energetics, with exchange by magnetic pressure gradients being negligible across all scales.  Magnetic energy is supplied by large-scale, $k L / (2 \pi) \lesssim 10$, kinetic energy loss due to turbulent fluid motions working against the magnetic tension force, as evidenced by negative values of $T_{\rm BKT}(k)$ on large scales.  The positive values of $T_{\rm BKT}(k)$ on intermediate scales peaking at $k L / (2 \pi) \sim 80$ indicate that magnetic energy is also being placed into intermediate-scale kinetic energy by the reversal of the process just described.  Note that the transfer function $T_{\rm BKT}(k)$ reveals that a significant amount of energy is being exchanged, but one cannot say on what scales it is distributed in the magnetic energy reservoir.  Presumably, some of this large-scale kinetic energy is transferred into large-scale magnetic energy.  The kinetic energy reservoir contributes a modest amount of intermediate/small-scale magnetic energy via work done on the magnetic field by fluid motions, as shown by positive values of $T_{\rm KBT}(k)$.  Most of the small-scale magnetic energy comes from turbulent transfer within the magnetic energy reservoir from large-to-small scales (i.e., $T_{\rm BBA}$ transitions from negative to positive values going from large-to-small scales).  Thus, Figure \ref{fig:Txbx_last} tells a story of a mechanism for ongoing large-scale magnetic energy production and a turbulent cascade from large-to-small scales within the magnetic energy reservoir.  This small-scale energy is exchanged forwards and backwards with the kinetic energy reservoir and is gradually dissipated, allowing the magnetic field to keep a relatively sustained value in the absence of a driven shear layer.

\begin{table}
\addtolength{\tabcolsep}{-3.25pt}
\centering
\begin{tabular}{c c c c c c c c}
\hline
\hline
Transfer Rate & $T_{\rm BKT}$ & $S_{\rm BKT}$ & $T_{\rm KBT}$ & $S_{\rm KBT}$ & $T_{\rm BKP}$ & $S_{\rm BKP}$ & $T_{\rm KBP}$ \\
 & (\%) & (\%) & (\%) & (\%) & (\%) & (\%) & (\%) \\
\hline
$\langle d E_{\rm M}^{+} / d t \rangle$ & 53.9 & 7.4 & 29.0 & 6.8 & 0.5 & 2.4 & 0.0 \\
$\langle d E_{\rm M}^{-} / d t \rangle$ & 69.5 & 0.6 & 11.7 & 0.0 & 5.0 & 4.6 & 8.6 \\
\hline
\end{tabular}
\caption{Percentage breakdowns of the contributions to the magnetic energy gain and loss rates by transfer function.  The {\it leftmost} column is the time-averaged magnetic energy transfer rate, where the superscripts denote $+$ for gain and $-$ for loss.  The {\it rightward} columns list the percentage contribution from each transfer function involved in magnetic energy exchange.  Notably, energy transfer mediated by turbulent motions (i.e., the $\mathbf{v}_{\rm t}$ component of $\mathbf{v}$) interacting with the magnetic tension force are the primary players in energy exchange with the magnetic energy reservoir.}
\label{tab:energytransfer}
\end{table}

\begin{figure*}
  \centering
  \includegraphics[angle=90, scale=0.6, trim=190 0 110 0, clip=true]{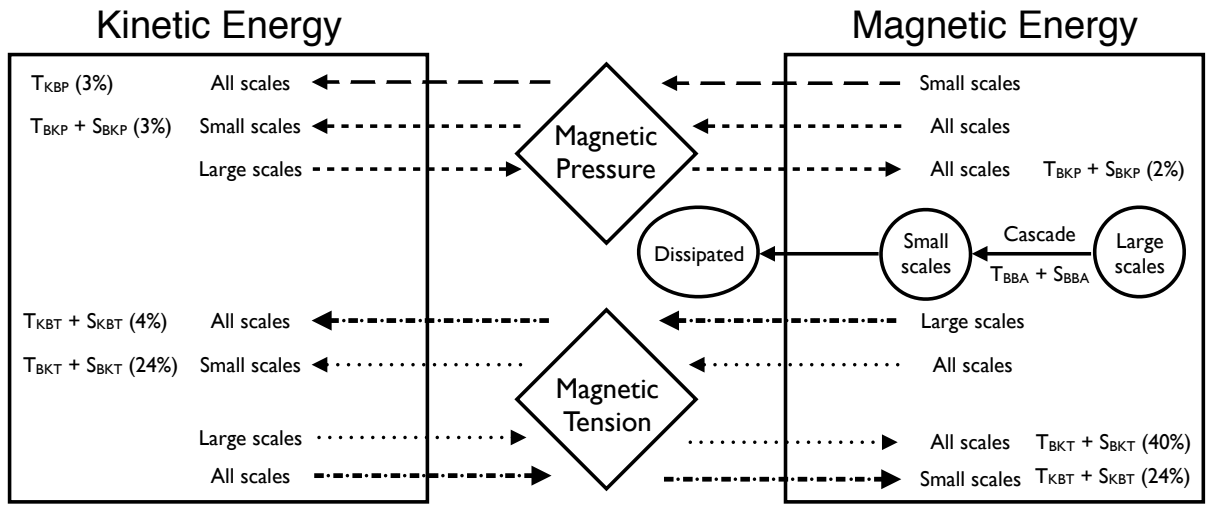}
  \caption{Diagram showing the contributions from transfer functions to the two-way energy exchange between the kinetic and magnetic energy reservoirs at late times.  Percentages in parentheses indicate the amount of energy exchange into that reservoir as described by the associated transfer functions, relative to the time-averaged total energy exchange rate, $\langle d E_{\rm M}^{\rm tot} \rangle$, over the time interval [$t_{\rm peak}$, $t_{\rm f}$].  The dominant scales (i.e., small, large, all) across which the energy transfer operates are indicated for each exchange path.  For instance, transfer of large-scale kinetic energy into the magnetic energy reservoir by twisting/stretching of magnetic field by fluid motions (i.e., $T_{\rm BKT} < 0$ and $S_{\rm BKT} < 0$) is responsible for 40\% of the total energy exchange.  The line styles are chosen to overlap with those of Figure \ref{fig:Txbx_last}.  Magnetic tension is the dominant transfer mechanism for exchanges into/out of the magnetic energy reservoir.  The kinetic/magnetic energy reservoir interactions result in a net magnetic energy gain rate.  This energy then cascades from large-to-small scales and is further exchanged forwards and backwards with the kinetic energy reservoir until it is ultimately dissipated.}
    \label{fig:energytransfer}
\end{figure*}

Finally, we perform an inventory of energy transfer operating in the KHI over the late-time turbulent decay stage from $t_{\rm peak}$ to $t_{\rm f}$ for the fiducial simulation 3M512.  Separately collecting the positive and negative contributions from each transfer function involved with exchange with the magnetic energy reservoir allows one to determine the total magnetic energy gain and loss rates due to energy exchanges,
\begin{align}
\frac{d E_{\rm M}^{+}}{d t} =&~ \int [ T_{\rm BKT}^{-}(k) + T_{\rm KBT}^{+}(k) + T_{\rm BKP}^{-}(k) + T_{\rm KBP}^{+}(k) + \nonumber \\
&~ S_{\rm BKT}^{-}(k) + S_{\rm KBT}^{+}(k) + S_{\rm BKP}^{-}(k) ] dk \label{eq:MEgain} \\
\frac{d E_{\rm M}^{-}}{d t} =&~ \int [ T_{\rm BKT}^{+}(k) + T_{\rm KBT}^{-}(k) + T_{\rm BKP}^{+}(k) + T_{\rm KBP}^{-}(k) + \nonumber \\
&~ S_{\rm BKT}^{+}(k) + S_{\rm KBT}^{-}(k) + S_{\rm BKP}^{+}(k) ] dk. \label{eq:MEloss}
\end{align}
The $\pm$ notation in the superscript indicates whether the positive (i.e., $\ge 0$) or negative (i.e., $< 0$) component of the transfer function should be taken. We find a time-averaged magnetic energy gain rate of $\langle d E_{\rm M}^{+} / d t \rangle = 4.4 \times 10^{-3}$ and loss rate of $\langle d E_{\rm M}^{-} / d t \rangle = -2.3 \times 10^{-3}$.  This gives a time-averaged net magnetic energy gain rate due to energy transfer with the magnetic energy reservoir of $\langle d E_{\rm M}^{\rm net} / d t \rangle = 2.1 \times 10^{-3}$, all in code units.  Note that the transfer functions describing the magnetic energy cascade (i.e., $T_{\rm BBA}(k)$ and $S_{\rm BBA}(k)$) are not included in this inventory because they cannot contribute to overall magnetic energy gain or loss.

Table \ref{tab:energytransfer} lists the relative contributions of each transfer function involved in magnetic energy gain and loss rates (see equations \ref{eq:MEgain} and \ref{eq:MEloss}) averaged from $t_{\rm peak}$ to $t_{\rm f}$.  Stretching and twisting of magnetic field lines by the turbulent velocity field (i.e., $T_{\rm BKT}(k)$ and $T_{\rm KBT}(k)$) is the dominant exchange mechanism at work during late times, accounting for 83\% and 81\% of energy transfer leading to magnetic energy gain and loss, respectively.  Magnetic pressure is a negligible contributing transfer mechanism for the subsonic and sub-Alf\'enic flows we consider.

To understand the two-way energy flow into/out of the magnetic energy reservoir, we consider the total time-averaged magnetic energy exchange rate, $\langle d E_{\rm M}^{\rm tot} / d t \rangle = \left| \langle d E_{\rm M}^{+} / d t \rangle \right| + \left| \langle d E_{\rm M}^{-} / d t \rangle \right|$, and construct a schematic diagram in Figure \ref{fig:energytransfer} that tracks the contributions of each transfer function to $\langle d E_{\rm M}^{\rm tot} / d t \rangle$.  Figure \ref{fig:energytransfer} illustrates that the kinetic energy reservoir interacts with the large-scale field and injects energy into the magnetic energy reservoir.  This energy cascades down to smaller scales and is exchanged backwards and forwards with the kinetic energy reservoir, before ultimately being dissipated.  The turbulent cascade from large-to-small scales (i.e., $T_{\rm BBA}(k)$) operates on 61\% of $\langle d E_{\rm M}^{+} / d t \rangle$, making the cascade within the magnetic energy reservoir an effective mechanism for breaking down magnetic structures.

Simultaneous with the magnetic energy reservoir exchange described by the transfer functions are net magnetic energy loss rates resulting from both the decaying nature of the MHD turbulence (i.e., $d E_{\rm M} / d t$) and numerical dissipation of magnetic energy (i.e., $D_{\rm M}$).  The time-averaged magnetic energy decay rate (i.e., left hand side of equation \ref{eq:transfunc_ME}) is $\langle d E_{\rm M} / d t \rangle = -1.6 \times 10^{-3}$.  The time-averaged numerical magnetic energy dissipation rate is $\langle D_{\rm M} \rangle = -3.0 \times 10^{-3}$, which is computed from equation \ref{eq:transfunc_ME} and integrating across all $k$.

\section{Dissipation}
\label{sec:dissip}
The extremely large Reynolds numbers that characterize astrophysical flows suggest that it is appropriate to carry out numerical simulations of these same flows in the inviscid flux-freezing regime, where explicit dissipation terms are omitted from the momentum and induction equations.  In nature, however, astrophysical flows have some small, but finite amount of viscosity and resistivity, which violates the assumption of an inviscid, flux-frozen flow.  In a turbulent flow, such as that considered here, it is these dissipation terms that mediate the dissipation of small-scale turbulent structures and conversion of magnetic and kinetic energy contained in these structures into thermal energy. When performing calculations in the inviscid, flux-freezing regime, simulators hope that the details of dissipation, which are provided by the algorithm, have little influence on large-to-intermediate scales.  If dissipation {\it does} influence these scales, simulators hope that the details of the numerical dissipation are sufficiently similar to physical dissipation such that the simulation remains an accurate representation of the physical system.  This non-trivial issue regarding the validity of relying on numerical dissipation to adequately capture the behaviour of physical dissipation in simulations is what we address in the present section.

\begin{figure}
  \includegraphics[width=84mm]{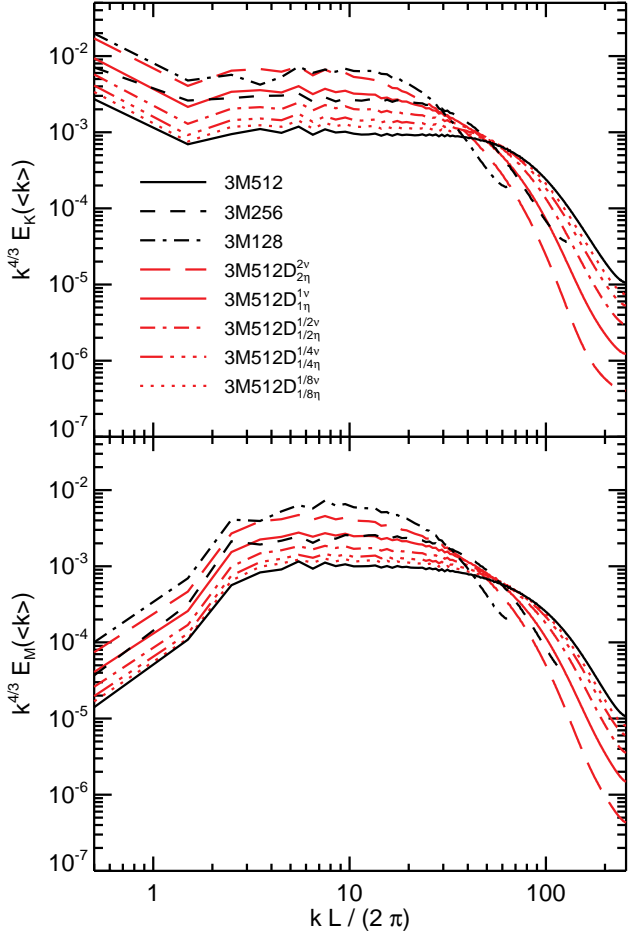}
  \caption{Time-averaged, one-dimensional, shell-averaged spectral energy densities for simulations with ({\it red lines}) and without ({\it black lines}) explicit dissipation.  Simulations with explicit dissipation introduced at saturation are: 3M512D$^{2 \nu}_{2 \eta}$ ({\it long-dashed lines}); 3M512D$^{1 \nu}_{1 \eta}$ ({\it solid lines}); 3M512D$^{1/2 \nu}_{1/2 \eta}$ ({\it dash-dot lines}); 3M512D$^{1/4 \nu}_{1/4 \eta}$ ({\it triple dot-dash lines}); 3M512D$^{1/8 \nu}_{1/8 \eta}$ ({\it dotted lines}). Simulations without explicit dissipation are: 3M512 ({\it solid lines}); 3M256 ({\it dashed lines}); 3M128 ({\it dotted lines}). Time averages are performed over [$t_{\rm peak}$, $t_{\rm f}$].  {\it Top panel}:  Kinetic energy power spectra, $E_{\rm K}(\langle k \rangle)$.  {\it Bottom panel}:  Magnetic energy power spectra, $E_{\rm M}(\langle k \rangle)$. Both spectral energy distributions are compensated by $k^{4/3}$. For all dissipation coefficients explored, power is depleted on small scales for the runs with explicit dissipation compared to the fiducial ideal MHD run.  Simulations 3M512$^{1/8 \nu}_{1/8 \eta}$ and 3M512$^{1 \nu}_{1 \eta}$, which incorporate explicit dissipation terms, provide close matches to the ideal MHD simulations 3M512 and 3M256, respectively.}
  \label{fig:powspec_3Dconv}
\end{figure}

\begin{figure}
  \includegraphics[width=84mm]{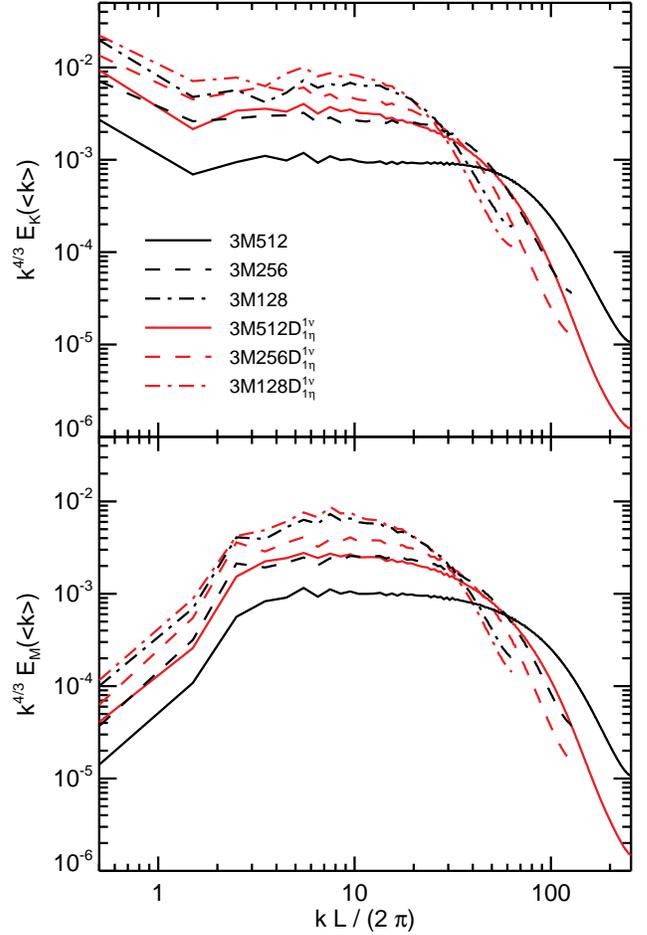}
  \caption{Time-averaged, one-dimensional, shell-averaged spectral energy densities for simulations with ({\it red lines}) and without ({\it black lines}) explicit dissipation.  Simulations with explicit dissipation introduced at saturation are: 3M512D$^{1 \nu}_{1 \eta}$ ({\it solid lines}); 3M256D$^{1 \nu}_{1 \eta}$ ({\it dashed lines}); 3M128D$^{1 \nu}_{1 \eta}$ ({\it dash-dot lines}).  Simulations without explicit dissipation are: 3M512 ({\it solid lines}); 3M256 ({\it dashed lines}); 3M128 ({\it dash-dot lines}). Time averages are performed over [$t_{\rm peak}$, $t_{\rm f}$].  {\it Top panel}:  Kinetic energy power spectra, $E_{\rm K}(\langle k \rangle)$.  {\it Bottom panel}:  Magnetic energy power spectra, $E_{\rm M}(\langle k \rangle)$. Both spectral energy distributions are compensated by $k^{4/3}$.  Numerical dissipation becomes a more dominant contributor to the total dissipation with decreasing numerical resolution.}
  \label{fig:powspec_3Ddissip}
\end{figure}

Figure \ref{fig:powspec_3Dconv} shows shell-averaged (see \S \ref{sec:specanal}) spectral energy densities obtained from simulations of decaying turbulence arising from the KHI with explicit dissipation added to the momentum and induction equation.  Specifically, we include the effects of kinematic shear viscosity and Ohmic resistivity.  The data of the convergence study presented in Figure \ref{fig:powspec_3Dres} are also shown on Figure \ref{fig:powspec_3Dconv} for comparison. The simulations with explicit dissipation were initialized from the fiducial ideal MHD simulation, 3M512, at $t=t_{\rm peak}$.  We found that kinematic shear viscosity and Ohmic resistivity coefficients, $\nu = 3.25 \times 10^{-6}$ and $\eta = 2.125 \times 10^{-6}$ ($Pm \equiv \nu / \eta = 1.53$, simulation 3M512D$^{1/8 \nu}_{1/8 \eta}$), produced a small, but non-negligible change in the magnetic and kinetic spectral energy densities over the time interval $t_{\rm peak}$ to $t_{\rm f}$ compared to the fiducial simulation. The dissipation coefficients were then increased by factors of two (at fixed magnetic Prandtl number, $Pm$, and initialized from $t = t_{\rm peak}$ of 3M512) until the magnetic and kinetic spectral energy densities provided a close match to those obtained from simulation 3M256.  This occurs for simulation 3M512D$^{1\nu}_{1\eta}$, where $\nu_{\rm fid} = 2.6 \times 10^{-5}$ and $\eta_{\rm fid} = 1.7 \times 10^{-5}$ (where subscript `fid' denotes that we treat these as our fiducial values), a factor of 8 increase over the dissipation coefficients found to match 3M512. This result suggests that the numerical dissipation present in the ideal simulations scales as $(\Delta x)^3$ rather than $(\Delta x)^2$ as would be expected for second-order convergence.  While the algorithms in \texttt{Athena} are overall second-order, the spatial reconstruction method used in the ideal simulations is third-order, perhaps suggesting that the numerical dissipation in the KHI problem is determined by the order of spatial reconstruction.  To demonstrate that this scaling holds generally for \texttt{Athena}-run KHI models would require an ensemble of 3D simulations including dissipation, which is beyond the scope of this work.

Figure \ref{fig:powspec_3Ddissip} examines the convergence of the simulations using the fiducial dissipation coefficients, where the data from the ideal MHD convergence study presented in Figure \ref{fig:powspec_3Dres} are included for comparison. At resolutions lower than $N = 512$ (i.e., $N=128,256$) we find that numerical dissipation plays an increasingly important role. In particular, there is a close correspondence across all scales between the simulations at $N = 128$ with (3M128D$^{1 \nu}_{1 \eta}$) and without (3M128) contributions from explicit dissipation, indicating that solutions at this resolution are dominated by effects due to numerical dissipation. The $N = 256$ case with the fiducial dissipation coefficients, 3M256D$^{1 \nu}_{1 \eta}$, matches the large-scale behaviour of 3M128 and the small-scale behaviour of 3M256.  As noted previously, the $N = 512$ case with the fiducial dissipation coefficients, 3M512D$^{1 \nu}_{1 \eta}$, provides a close match to results obtained for 3M256 at all scales.  The primary difference from 3M256 is a small power deficit for 3M512D$^{1 \nu}_{1 \eta}$ at scales around the dissipation scale\footnote{The dissipation scale refers to the approximate turn-over scale where the spectral energy distribution transitions from a power-law inertial range to the steep decline at small scales.}, $k L / (2 \pi) \sim 30$, due to power being transferred over to smaller scales, $k L / (2 \pi) \sim 100$, where there is a slight power excess.  We regard the simulation using the fiducial dissipation coefficients as being converged at $N = 512$ for two reasons.  First, we already demonstrated that simulations conducted in ideal MHD are converged at this resolution (see \S \ref{sec:converge}), implying that numerical dissipation plays a small role in simulations at this resolution. Second, for simulations incorporating physical dissipation, convergence implies that the dissipation scale is resolved. As elucidated above, the location of the dissipation scale for simulation 3M512D$^{1\nu}_{1\eta}$ moved to smaller $k$ (i.e., larger physical scales) compared to the ideal simulation at the same numerical resolution, 3M512. This implies that the location of the dissipation scale is determined by the dissipation terms themselves rather than numerical effects.  Therefore, we can conclude that the dissipation scale associated with $\nu_{\rm fid}, \eta_{\rm fid}$ is resolved at $N = 512$.

\begin{figure}
  \includegraphics[width=84mm]{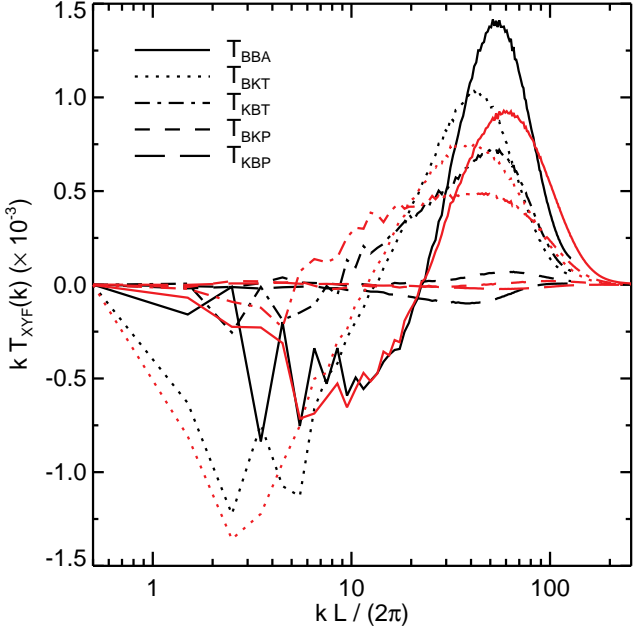}
  \caption{Time-averaged, one-dimensional spectral energy transfer functions associated with energy transfer into/out of the magnetic energy reservoir for the ideal MHD run 3M256 ({\it black lines}) and the simulation 3M512D$^{1\nu}_{1\eta}$ ({\it red lines}) where explicit dissipation was introduced at saturation with fiducial dissipation coefficients.  The time-averaging interval and choice of line styles are consistent with those of Figure \ref{fig:Txbx_last}.  The dominant transfer functions in energy exchange ($T_{\rm BBA}(k)$ and $T_{\rm BKT}(k)$) are generally well-matched between these two runs at scales larger than the dissipation scale (i.e., $k \lesssim 30$), suggesting that the physics of energy transfer in MHD turbulence is robust to the effects of numerical dissipation.}
  \label{fig:Tfunc_dissip}
\end{figure}

\begin{figure}
  \centering
  \includegraphics[width=\columnwidth]{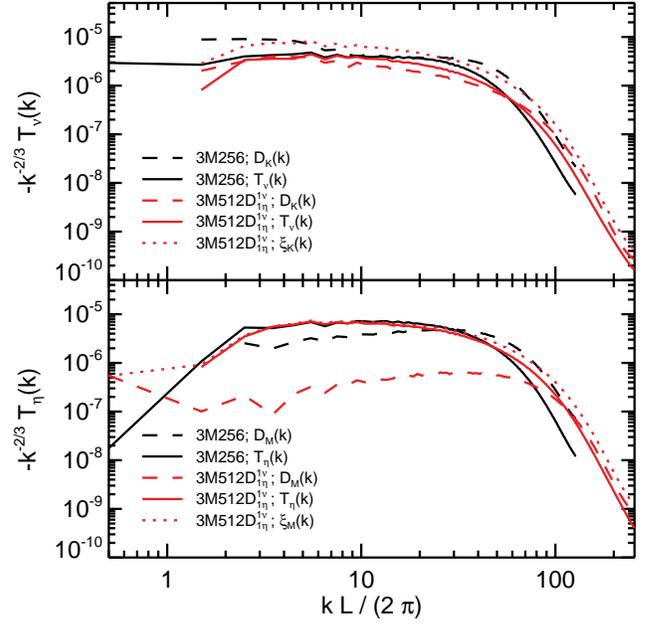}
  \caption{Time-averaged, one-dimensional dissipation transfer functions for the 3M256 ({\it black lines}) and 3M512D$^{1 \nu}_{1 \eta}$ ({\it red lines}) simulations over the interval, [$t_{\rm peak}$, $t_{\rm f}$] and compensated by $-k^{-2/3}$.  {\it Top panel}: Kinetic energy dissipation transfer functions and numerical dissipation terms.  {\it Bottom panel}: Magnetic energy dissipation transfer functions and numerical dissipation terms.  See the legends within each panel for the specific quantities being shown.}
  \label{fig:dissip}
\end{figure}

With these arguments in mind, Figure \ref{fig:Tfunc_dissip} compares transfer functions associated with energy exchange with the magnetic energy reservoir for simulations 3M512D$^{1\nu}_{1\eta}$ and 3M256, time-averaged over the interval [$t_{\rm peak}, t_{\rm f}$].  At large spatial scales, $k \lesssim 5$, the transfer functions for 3M512D$^{1\nu}_{1\eta}$ and 3M256 are well-matched. At intermediate scales, $5 \lesssim k \lesssim 30$, the transfer functions are well-matched for transfer from magnetic energy to magnetic energy through advection, $T_{\rm BBA}(k)$, and from magnetic energy to kinetic energy through tension forces, $T_{\rm BKT}(k)$. By contrast, we see greater transfer from kinetic to magnetic energy through tension, $T_{\rm KBT}(k)$, at these intermediate scales for 3M512D$^{1\nu}_{1\eta}$ than for 3M256.  At small scales, $k \gtrsim 30$, we see that peaks in the transfer functions are shifted to smaller scales in 3M512D$^{1\nu}_{1\eta}$, as compared to 3M256, as a consequence of the higher numerical resolution in this simulation.  Finally, at all scales, the effect of dissipation is to reduce the (already small) contribution of energy transfer through compressive motions, $T_{\rm KBP}(k)$ and $T_{\rm BKP}(k)$. Overall, these results demonstrate the robustness of the physics of energy transfer within decaying MHD turbulence to the effects of numerical dissipation at scales larger than the dissipation scale.

The transfer functions associated with explicit dissipation take the form (see Appendix \ref{append:transfunc}),
\begin{align}
T_{\nu}(k) &= \frac{\nu}{2} \left( \widehat{\left[ \rho \mathbf{v} \right]} \cdot \widehat{\left[ \frac{1}{\rho} \left( \nabla \cdot \bm{\tau} \right) \right]^{\ast}(k)} + \widehat{\mathbf{v}} \cdot \widehat{\left( \nabla \cdot \bm{\tau} \right)}^{\ast}(k) \right) \label{eq:Tnu} \\
T_{\eta}(k) &= \eta \left( \widehat{\mathbf{B}}(k) \cdot \left[ \widehat{\nabla^{2} \mathbf{B}} \right]^{\ast}(k) \right). \label{eq:Teta}
\end{align}
These are plotted in Figure \ref{fig:dissip} for simulations 3M256 and 3M512D$^{1\nu}_{1\eta}$.  Note that for ideal (i.e., inviscid, $\nu = 0$ and non-resistive, $\eta = 0$) MHD simulations, such as 3M256, the physical dissipation transfer functions $T_{\nu}(k)$ and $T_{\eta}(k)$ do not contribute to the overall energy transfer inventory, but are instead computed for the sake of establishing `effective' quantities.  The effective dissipation transfer function data of 3M256 adopt $\nu_{\rm fid}$ and $\eta_{\rm fid}$ to enable comparison with $T_{\nu}(k)$ and $T_{\eta}(k)$ from 3M512D$^{1 \nu}_{1 \eta}$.  Also shown in Figure \ref{fig:dissip} for both simulations are the quantities $D_{\rm K}(k)$ and $D_{\rm M}(k)$.  These numerical dissipation rates are derived by calculating the residual between the terms in Equations \ref{eq:transfunc_KE} and \ref{eq:transfunc_ME} (including $T_{\nu}(k)$ and $T_{\eta}(k)$ for 3M512D$^{1\nu}_{1\eta}$) and the total dissipation rates ($\xi_{\rm K}(k)$ and $\xi_{\rm M}(k)$) for simulation 3M512D$^{1\nu}_{1\eta}$.  The total kinetic and magnetic energy dissipation rates are expressed as,
\begin{align}
\xi_{\rm K}(k) &= T_{\nu}(k) + D_{\rm K}(k) \label{eq:xiK} \\
\xi_{\rm M}(k) &= T_{\eta}(k) + D_{\rm M}(k) \label{eq:xiM}.
\end{align}
Figure \ref{fig:dissip} shows that the spectral distribution of $T_{\nu}(k)$, $T_{\eta}(k)$ is very similar between the two simulations for scales larger than the dissipation scale and that the spectral distribution of numerical dissipation in 3M256 is very close to that expected from physical dissipation close to the grid scale (i.e., small scales).  A further point comes from comparing physical and numerical dissipation in simulation 3M512D$^{1\nu}_{1\eta}$.  For this simulation, physical dissipation, $T_{\eta}(k)$, dominates over numerical dissipation, $D_{\rm M}(k)$, for the magnetic energy dissipation terms by a factor $\sim 100$ for $2 \lesssim k L / (2 \pi) \lesssim 30$.  However, the same is not true for the kinetic energy dissipation terms, where physical dissipation, $T_{\nu}(k)$, and numerical dissipation, $D_{\rm K}(k)$, are relatively comparable to within a factor of $\sim 2$ over this range of scales.  Similar levels of numerical and physical kinetic energy dissipation could be due to the computation of derivatives that are required for the viscous stress tensor, $\bm{\tau}$, and the associated divergence, $\nabla \cdot \bm{\tau}$. The same considerations do not apply for the addition of Ohmic diffusion to the induction equation due to the use of the CT algorithm for these terms, which may explain why physical magnetic energy dissipation greatly exceeds numerical magnetic energy dissipation on scales larger than the dissipation scale.

\section{Summary and Discussion}
\label{sec:sumdisc}
We performed a suite of 2D and 3D simulations of the KHI in the weakly magnetized, subsonic regime with a non-driven shear layer, focusing on the results of a high-resolution 3D MHD simulation.  The problem setup, though simple and straightforward, was scrutinized in detail, paying particular attention to dimensionality (2D {\it versus} 3D), convergence, and properly resolving the shear layer in order to make a convincing argument for the physical nature of the KHI development beyond the linear growth.  After establishing the basic evolution of energetics using volume-averaged energies and time-averaged energy power spectra, we took advantage of the energy conserving nature of \texttt{Athena} to investigate the spectral structure of the KHI development into MHD turbulence using spectral energy transfer function analysis.  We then extended this analysis to characterize both numerical and physical dissipation in \texttt{Athena}.  Here, we discuss our results.

Two-dimensional MHD simulations of the KHI \citep[e.g.,][]{Franketal1996, Jones1997, Jeongetal2000, BucciantiniDelZanna2006} are attractive due to their ability to achieve high resolution relative to their 3D counterparts.  However, a demonstration of convergence of the resulting turbulent flow is required to justify 2D studies of MHD turbulence arising from the KHI.  We observe well-converged solutions of the initial growth of the 2D KHI at the moderate resolution $N = 512$, which justifies the linear growth stage of the 2D KHI as a highly reliable, robust test for code verification as suggested by \citet{McNallyLyraPassy2012}.  However, the saturated state and level of magnetic energy sustainment fails to converge even out to the extremely high 2D resolution $N = 16,384$, as evidenced by both the time evolution of the volume-averaged magnetic field strength (see Figure \ref{fig:Bamp2D}) and the changing shape of spectral energy densities with resolution (see Figure \ref{fig:powspec_2Dres}).  In stark contrast to the 2D case, 3D simulations of the KHI reliably converge at a resolution $N = 512$ over the full-course of evolution out to the turbulent and decaying stages.

Time evolutions of volume-averaged energetics and slices of the simulation volume reveal a decline in kinetic energy and growth of magnetic energy to a saturated level, at which time the shear layers are almost completely disrupted.  The subsequent evolution leads to turbulence with a sustained, but gradually decaying, magnetic field.  This general evolution is also observed in relativistic MHD simulations of the KHI when the driving mechanism is switched off \citep{BucciantiniDelZanna2006, ZhangMacFadyenWang2009, ZrakeMacFadyen2011}.  These studies adopt either a discontinuous shear layer, use a Riemann solver of type HLLE, or both.  We find that the decline in kinetic energy during the non-linear growth and generation of a sustained magnetic field is robust to the details of the initial setup and Riemann solver used (see Appendix \ref{append:numerical_issues}).  We confirm the results of the relativistic MHD study of the KHI of \citet{BeckwithStone2011} in the Newtonian regime using a linearized Riemann solver.  While the generic result of the appearance of a saturated state is unaffected, we caution against using a setup with an unresolved interface and/or the HLLE Riemann solver for quantitative studies of the KHI.

\begin{figure}
  \centering
  \includegraphics[width=84mm]{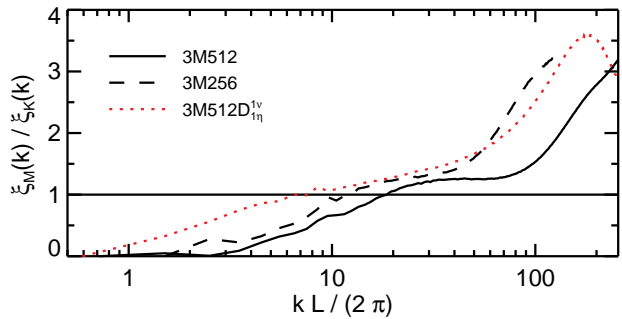}
  \caption{Ratio of total (i.e., numerical + physical, if applicable) magnetic-to-kinetic energy dissipation rates for simulations 3M512 ({\it solid line}), 3M256 ({\it dashed line}), and 3M512D$^{1 \nu}_{1 \eta}$ ({\it dotted line}), time-averaged over the interval [$t_{\rm peak}$, $t_{\rm f}$].  The {\it horizontal solid line} marks the equipartition between the total magnetic energy dissipation rate, $\xi_{\rm M}(k)$, and the total kinetic energy dissipation rate, $\xi_{\rm K}(k)$.  Magnetic energy dissipation exceeds kinetic energy dissipation for intermediate-to-small scales (i.e., $k \gtrsim 10$).}
  \label{fig:dissip_DmDk}
\end{figure}

The spectral distributions of kinetic and magnetic energy for 3D KHI simulations at late times follow an approximate $k^{-4/3}$ power-law on intermediate scales, $5 \lesssim k L / (2 \pi) \lesssim 30$, remaining unaltered for all resolutions considered (see Figure \ref{fig:powspec_3Dres}).  A spectral slope $\propto k^{-4/3}$ over intermediate scales also appeared in the strong-field driven supersonic MHD turbulence studies of \citet{LemasterStone2009} for 3D resolutions of $N = 512$ and $N = 1024$.  The effect of increasing numerical resolution is to move the dissipation scale to smaller scales.  The magnetic-to-kinetic energy spectral equipartition point shifts to larger scales throughout the simulation evolution (see Figure \ref{fig:powspec_allE}).  Performing a study of relativistic, ideal MHD turbulence arising from the KHI, \citet{ZhangMacFadyenWang2009} claim that this observed evolution of the $E_{\rm M}(k) / E_{\rm K}(k)$ equipartition point indicates that the kinematic viscous dissipation is more efficient than the magnetic resistive dissipation.  However, this conjecture was not based on a direct study of dissipation.  Figure \ref{fig:dissip_DmDk} shows the ratio of total magnetic-to-kinetic energy dissipation rates in the turbulent regime for simulations with (3M512D$^{1 \nu}_{1 \eta}$) and without (3M256, 3M512) explicit dissipation included.  Figure \ref{fig:dissip_DmDk} demonstrates that magnetic energy dissipation actually exceeds kinetic energy dissipation across the majority of scales, $k \gtrsim 10$.  Therefore, the shift in the $E_{\rm M}(k) / E_{\rm K}(k)$ equipartition point in Figure \ref{fig:powspec_allE} is instead a consequence of the exchange of large-scale kinetic energy into the magnetic energy reservoir mediated by turbulent motions acting against magnetic tension (i.e., fluid motions twisting/stretching magnetic field lines).  This is evidenced by the dominating negative values of the transfer function $T_{\rm BKT}(k)$ in Figures \ref{fig:Txbx} and \ref{fig:Txbx_last}.  Therefore, large-scale kinetic energy loss to the magnetic energy reservoir, rather than competing dissipation rates, is the true mechanism behind the shift in the $E_{\rm M}(k) / E_{\rm K}(k)$ equipartition point to large scales as the KHI evolves non-linearly.   Transfer function analysis resolved this ambiguity and this example illustrates that the transfer function diagnostic is a powerful tool for studying how energy is transferred across scales and forms.

Spectral energy transfer analysis allows for both the scale-by-scale quantification of energy transfer between reservoirs and identification of the mechanism responsible for the energy exchange.  This information is inaccessible from power spectra alone.  As the KHI develops to a saturated state, the growth of magnetic energy is dominated by the magnetic tension force interacting with turbulent motions and an inverse cascade is observed.  This means that magnetic energy is initially concentrated on small scales and then evolves to a spectrum dominated on large scales (see Figure \ref{fig:Txbx}).  At late times following saturation when the fluid is in a decaying turbulent state, we find no evidence for dynamo operation for a single fluid treatment.  This is contrary to claims from simulations of decaying turbulence arising from relativistic MHD KHI studies \citet{ZhangMacFadyenWang2009}.  Kinetic energy contained in turbulent fluid motions is transferred to magnetic energy, primarily mediated by interactions with the magnetic tension force, and a turbulent cascade from large-to-small scales operates within the magnetic energy reservoir.  This small-scale magnetic energy is interchanged forwards and backwards with the kinetic energy reservoir and is eventually dissipated, allowing the magnetic energy to decay.  For the subsonic and sub-Alfv\'enic relative flow considered in this work, compressible effects are of ancillary importance in energy transfer.

By their nature, numerical simulations exhibit dissipative behaviour due to finite numerical resolution.  Even in instances where physical dissipation terms are explicitly included in the solution of the MHD conservation equations, numerical dissipation is still present at some level.  We found that the most important effect of increasing numerical resolution for ideal MHD simulations was to move the dissipation scale to progressively smaller scales.  While energy dissipation in ideal MHD simulations occurs preferentially on the grid scale, physical dissipation should act across all scales.  Therefore, determining the extent to which numerical dissipation affects MHD turbulence when physical dissipation is present is nontrivial.  We found that when the numerical resolution was held fixed, the location of the dissipation scale moves to larger spatial scales when physical dissipation is incorporated (3M512D$^{1 \nu}_{1 \eta}$) compared to the corresponding ideal MHD simulation (3M512).  This result indicates that it is the dissipation terms that determine the dissipation scale, rather than numerical effects.  The physical dissipation scale is considered to be resolved when the dissipation scale (i.e., the turnover in the power spectrum at large $k$) moves to larger spatial scales than in the case without explicit dissipation terms included.  In this sense, the effective resolution of the simulation, by which we mean the location of the dissipation scale, is reduced by construction.  Furthermore, when physical dissipation is introduced, the magnitude of numerical dissipation is diminished and the spectral character of the transfer functions (i.e., general shapes and relative proportions) involved in exchange with the magnetic energy reservoir are well-matched to their ideal MHD counterparts.  These observations indicate the robustness of the physics of energy transfer in decaying MHD turbulence to the effects of numerical dissipation, at least for scales larger than the dissipation scale where numerical effects do not dominate.

\section{Conclusions}
\label{sec:conc}
We list our conclusions here followed by some astrophysical implications of this work.

\begin{enumerate}[label={\bfseries $\bullet$},leftmargin=*]
\item 3D KHI simulations converge --- in the virtual meaning (see \S \ref{sec:converge}) --- across all stages of evolution.  The main effect of further increasing numerical resolution is to push the numerical dissipation scale to smaller spatial scales without changing the shape of the power spectrum.

\item For subsonic, weakly magnetized, decaying turbulence arising from the non-driven KHI, the spectral distributions of kinetic and magnetic energy for 3D simulations follow an approximate $k^{-4/3}$ power-law on intermediate scales.

\item Spectral energy transfer function analysis is a powerful diagnostic for quantifying energetics and dissipation in MHD turbulence.

\item At late times corresponding to decaying MHD turbulence, energy is injected into the magnetic reservoir as a result of kinetic energy interactions with the large-scale magnetic field.  This magnetic energy turbulently cascades down to smaller scales and is exchanged backwards and forwards with the kinetic energy reservoir, before ultimately being dissipated.

\item Incorporating explicit dissipation terms reduces the importance of numerical dissipation and moves the dissipation scale to larger spatial scales.  For the levels of physical dissipation considered, introducing dissipation terms does not grossly alter the overall shape of the kinetic and magnetic energy power spectra.

\item The nature of numerical dissipation does not affect the physics of energy transfer within decaying MHD turbulence at scales larger than the dissipation scale, as evidenced by comparing the relative strengths of the transfer functions and dissipation rates.
\end{enumerate}

Our investigation of the subsonic KHI in the weak magnetic field limit and the generalized spectral energy transfer function techniques we exploit serve as a launching point for future studies of MHD turbulence and the extension to more targeted astrophysical applications of the KHI.

In addition to serving as a direct examination of KHI physics, this work provides a valuable baseline for investigations of shear layers in astrophysical systems also subject to the family of current-driven instabilities (CDI).  Such systems could potentially feature either sharp shear layers, such as those explored by \citet{BatyKeppens2002} or \citet{MizunoHardeeNishikawa2011}, or more gradual profiles, such as those examined analytically by \citet{NalewajkoBegelman2012}.  Regardless of the details, it should be possible to compare growth rates of systems unstable to the KHI and CDI to both analytic estimates of linear growth and those rates measured empirically in this work.  This will enable differentiation between CDI \citep{ONeill2012} and KHI contributions (this work) to energy evolution in these systems.  Furthermore, one could compare the non-linear evolution of turbulence examined here to similar turbulence that develops in joint KHI/CDI systems to determine how turbulent spectra, energy partitioning, and saturation levels differ between the two scenarios.

The transfer function machinery developed and used here can be applied to other astrophysically relevant systems.  In particular, local simulations of magnetized accretion discs in the ``mesoscale" regime (i.e., scales much larger than a vertical scale height but much less than the disk radius) by \cite{Simon2012} show that as larger disc scales are captured within the domain, turbulence driven by the magnetorotational instability \cite[MRI;][]{BalbusHawley1998} develops structure on these larger scales at the expense of small scale structure.  This behaviour is indicative of either an inverse cascade of energy or direct communication between small scales and large scales. In either case, applying our transfer function analysis to these mesoscale simulations will lead to a better understanding of energy flow in MRI turbulent disks.

While astrophysical scenarios often lend themselves nicely to powerful computational studies, various obstacles (e.g., numerical convergence, multitude of important physical processes, wide range in physical and temporal scales) force numericists to omit certain physics.  When restricted to the ideal (i.e., inviscid and non-resistive) MHD limit, one often conjectures that numerical dissipation behaves sufficiently similarly to physical dissipation, even in situations where dissipation may be an important physical process for the problem at hand.  For instance, dissipation of turbulence arising from the MRI is an important problem in compact object accretion disc physics, yet these studies are commonly performed in the ideal MHD limit.  As an example, attempts to model the effect of the vertical dissipation profile on the emergent accretion disc spectrum are very important for understanding observations of X-ray binaries \citep{Turner2004, Hirose2006, Blaes2006}.   A reasonable question to ask is whether numerical dissipation leads to unwanted numerical artifacts in the absence of physical dissipation.  Our work demonstrates that the details of numerical dissipation do {\it not} affect the physics of KHI-produced MHD turbulence on scales larger than the dissipation scale.  Therefore, studies of ideal MHD turbulence conducted with codes comparable to \texttt{Athena} are not plagued by numerical effects due to the nature of numerical dissipation.

\section*{Acknowledgements}
The authors thank the anonymous referee for her/his constructive comments and suggestions, which improved this paper.  GS thanks the National Science Foundation (NSF) for support through the Graduate Research Fellowship Program.  The authors acknowledge grant support through the NSF (AST-0807471; AST-0907872), NASA (NNX09AB90G; NNX09AG02G; NNX11AE12G; NNX12AE33G), University of Colorado Boulder (subcontract \#000115783), and support from Tech-X Corporation.  KB acknowledges useful discussions and advice from James Stone, Colin McNally, and Axel Brandenburg.  KB thanks the organisers and attendees of the ``NORDITA Astrophysics Code Comparison Workshop'' held in Stockholm, Sweden in August 2012 for stimulating discussions that helped in the formulation of this work.  This work used the \texttt{JANUS} supercomputer (account number UCB00000022), which is supported by the NSF (award number CNS-0821794) and the University of Colorado Boulder.  The \texttt{JANUS} supercomputer is a joint effort of the University of Colorado Boulder, the University of Colorado Denver, and the National Center for Atmospheric Research.

\bibliographystyle{mn2e}
\bibliography{ms}
\label{lastpage}

\appendix
\section{Numerical Issues}
\label{append:numerical_issues}
Here, we address some of the numerical issues in the KHI simulations with the \texttt{Athena} code in order to justify our particular setup.

\subsection{Riemann Solvers}
\label{append:riemann}
\begin{figure}
  \centering
  \includegraphics[width=84mm]{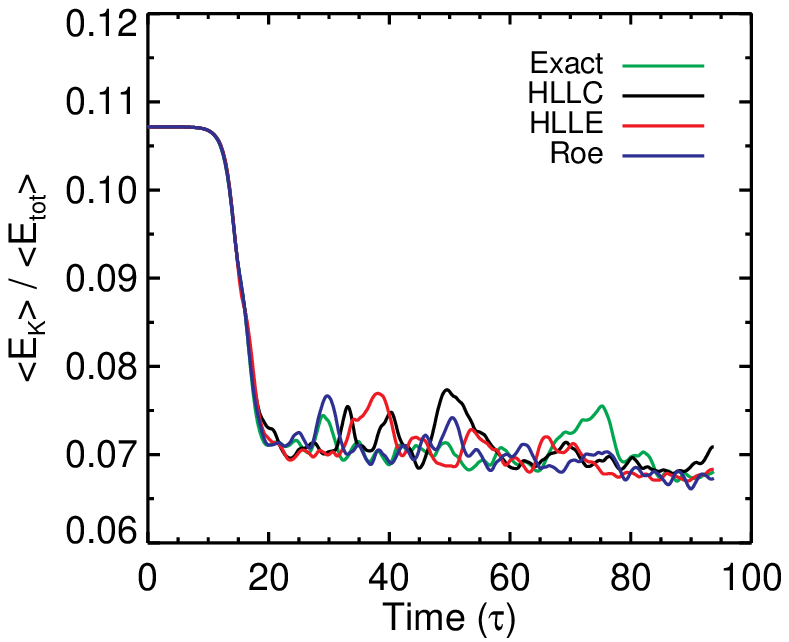}
  \includegraphics[width=84mm]{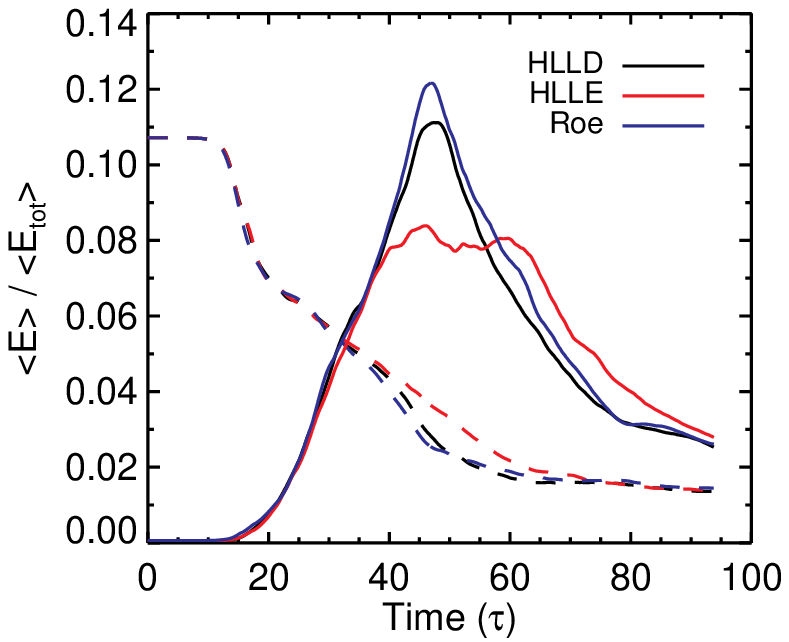}
  \caption{Comparison of Riemann solver performance in 2D KHI simulations with resolution $(N_{x} \times N_{y}) = (2048 \times 2048)$ computed with the \texttt{Athena} code.  {\it Top panel}: Time evolution of volume-averaged kinetic energy, $\left<E_{\rm K}\right>$, relative to the volume-averaged total energy in the computational box, $\left<E_{\rm tot}\right>$, for hydrodynamic simulations performed with the exact ({\it green line}), HLLC ({\it black line}), HLLE ({\it red line}), and Roe ({\it blue line}) Riemann solvers.  The evolution of $\left<E_{\rm K}\right>$ is essentially independent of the chosen solver in hydrodynamical simulations.  {\it Bottom panel}: Time evolution of volume-averaged magnetic energy, $\left<E_{\rm M}\right>$, multiplied by a factor of 5 ({\it solid lines}) and volume-averaged kinetic energy, $\left<E_{\rm K}\right>$ ({\it dashed lines}), each relative to $\left<E_{\rm tot}\right>$, for MHD simulations performed with the HLLD ({\it black line}), HLLE ({\it red line}), and Roe ({\it blue line}) Riemann solvers.  The saturation level of $\left<E_{\rm M}\right>$ for the run adopting the HLLE solver is $\sim 30$\% below that of the runs that used the HLLD and Roe solvers.}
  \label{fig:riemann}
\end{figure}

The available suite of Riemann solvers implemented in \texttt{Athena} for approximating physical fluxes across cell interfaces are the Roe, HLLD, and HLLE solvers for magnetohydrodynamics (MHD) and the exact, Roe, HLLC, and HLLE solvers for hydrodynamics (for descriptions of Riemann solvers, see \citealt{Toro1999} and \citealt{Leveque2002}).  Waves traveling between grid cells are dispersive and wave propagation can be visualized as a Riemann `fan', with the fastest-moving `leftward' and `rightward' waves defining the fan edges and an ensemble of intermediate-speed waves composing the fan.  The HLLD and HLLC solvers consider many intermediate-speed waves when computing fluxes, while the HLLE solver omits these waves and only accounts for the fastest waves propagating in each direction.  Unlike the HLL-- family of Riemann solvers, the Roe solver constructs the exact solution to a linearized form of the equations at the cell interfaces.

Figure \ref{fig:riemann} shows the comparison of Riemann solver performance in \texttt{Athena} for the two-dimensional KHI in hydrodynamics and MHD.  Note that the simulations shown in Figure \ref{fig:riemann} are equivalent to the 2H2048 and 2M2048 runs, except with different choices for the Riemann solver --- these additional simulations are not listed in Table \ref{tab:simlist}.  While all solvers capture similar linear growth rates as measured from the magnetic energy evolution, the HLLE solver diverges from Roe and HLLD during the onset of the non-linear evolution.  Specifically, magnetic energy in the HLLE run saturates at a level approximately $30 \%$ less than that of the Roe solver and spends a much longer time at a saturated state prior to entering the decay phase of the instability.  Such diffusive behavior in the HLLE solver has been reported in other contexts \citep{MignoneUglianoBodo2009, BeckwithStone2011,ONeill2012}, all of which suggest that previous investigations of the magnetized KHI that rely on the HLLE solver \citep[e.g.,][]{BucciantiniDelZanna2006, ZhangMacFadyenWang2009} may suffer from similar effects.  In our KHI simulations, we use the HLLC and HLLD solvers exclusively for hydrodynamics and MHD, respectively, with the \texttt{hllallwave} configure option turned on to include the full interpolated Riemann fan.

\subsection{Linear Growth}
\label{append:lingrow}
To demonstrate that our computational setup indeed produces sensible linear growth of the KHI, we compared the development of a simple, equal density realization of the KHI simulated with \texttt{Athena} to estimates of linear growth provided in \citet{Chandrasekhar1961} and \citet{MiuraPritchett1982}.  The expression in \citet{Chandrasekhar1961} describes the growth of an infinitesimally sharp shear layer in an incompressible, weakly magnetized fluid as $\Gamma \sim kU_0$ (using our notation), which corresponds to a value of $\Gamma \sim 3$ in our code units.  The growth rates in \citet{MiuraPritchett1982} are more applicable to our setup in that they incorporate a finite-width shear layer and compressibility, but unfortunately rely on the approximation that the modes are short in wavelength compared to the box size, which is not satisfied for our $k=2\pi/L$ perturbations.  The maximum growth rate from \citet{MiuraPritchett1982} most appropriate for our setup is $\Gamma \sim 7$, which is considerably faster than that of \citet{Chandrasekhar1961} because it occurs on a much smaller physical scale.  Empirically, our fastest growth rates are measured to be $\Gamma \sim 5$, which falls comfortably between the two estimates.  Furthermore, when we conducted KHI test cases featuring perturbations considerably smaller in scale than $L$, we found growth rates more comparable to those in \citet{MiuraPritchett1982}.  We therefore conclude that the linear development of the KHI in our simulations is consistent with theoretical expectations for the instability.

\subsection{Discontinuous {\it versus} Resolved Shear Layers}
\label{append:jump}
\begin{figure}
  \centering
  \includegraphics[width=84mm]{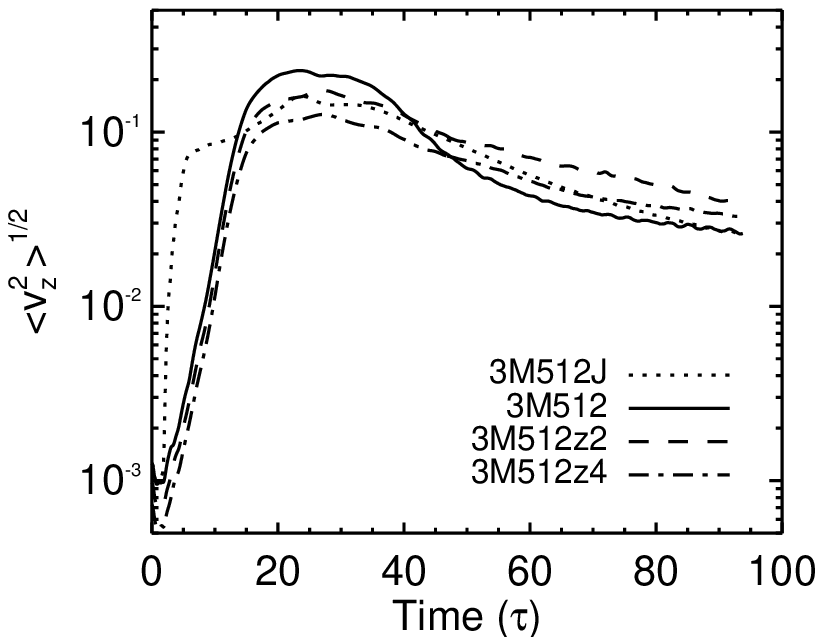}
  \caption{Volume-averaged rms velocity transverse to the shear layers for the resolved shearing model 3M512 ({\it solid line}), the discontinuous shearing model 3M512J ({\it dotted line}), and the extended domain models 3M512z2 ({\it dashed line}) and 3M512z4 ({\it dash-dot line}).  The linear growth rate for the discontinuous profile is dissimilar from those with resolved shear profiles.  Simulations with an extended $z$-domain exhibit different behaviour from the smaller box fiducial run during the early stages of the  non-linear decay phase of the instability.}
  \label{fig:Vz_rms_zbox_jump}
\end{figure}

An inadequately resolved shear interface may result in the accumulation of numerical truncation error, causing unphysical realizations of the subsequent evolution.  To quantify the degree to which the energetics are affected by the presence of an unresolved shear layer, we repeated the 3M512 simulation with the hyperbolic tangent interfaces for velocity and density replaced by jump discontinuities,
\begin{align}
  v_{y}(z) &= \left \{
    \begin{array}{lr}
      U_{0}, & \left| z \right| \ge z_{0} \\
      -U_{0}, & \left| z \right| < z_{0}
    \end{array}
  \right. \\
  \rho(z) &= \left \{
    \begin{array}{lr}
      1, & \left| z \right| \ge z_{0} \\
      2, & \left| z \right| < z_{0}
    \end{array}
  \right.
\end{align}
We refer to this simulation as 3M512J, where the J refers to the jump discontinuities in velocity and density across the shear interfaces.  Figure \ref{fig:Vz_rms_zbox_jump} compares the time evolution of the volume-averaged root mean square (rms) velocity transverse to the shear layers, $\langle v_{z}^{2} \rangle^{1/2}$, for the cases of resolved ({\it solid line}) and discontinuous ({\it dotted line}) interfaces.  The initial onset of instability for 3M512J occurs sooner than in 3M512 because the accumulation of truncation errors at the interface permits perturbations at smaller scales (i.e., faster growth rates) than would be available for a finite-width shear layer.  Despite the triggering of the KHI from an unresolved interface, the ultimate saturation and late-time evolution of 3M512J remains similar to that of 3M512.

\subsection{Extending the Domain Transverse to the Shear Layer}
\label{append:zbox}
The choice of periodic boundary conditions was motivated by its ease of implementation and its attractive consequence of energy conservation within the domain.  As the KHI evolves to the non-linear regime, propagating waves and fluid that exit through one boundary will re-enter through the opposite boundary and interact with the flow.  A potential concern is that cross-boundary interactions may substantially affect the evolution of the flow and produce an outcome driven by numerical, rather than physical, processes.

Figure \ref{fig:Vz_rms_zbox_jump} compares the evolution of the volume-averaged rms velocity transverse to the shear layers in model 3M512 with those of models 3M512z2 and 3M512z4, for which the $z$-domain is extended by a factor of two and four, respectively.  For the 3M512z4 simulation, the $z$-domain is so far extended that the turbulent regions are isolated in $z$ and the turbulence never crosses the $z$-boundaries.  Figure \ref{fig:Vz_rms_zbox_jump} shows that extending the domain transverse to the shear layer does not substantially affect the linear growth or the peak $\langle v_{z}^{2} \rangle^{1/2}$ amplitude that is achieved.  A difference arises only when the KHI is well into the non-linear decay phase.  At this point, the flows are both very turbulent, but $\langle v_{z}^{2} \rangle^{1/2}$ in the smallest domain decreases more rapidly after the peak amplitude than it does in the larger domains.  After this, however, the two decay rates become approximately parallel, suggesting that cross-boundary interactions do not grossly affect the asymptotic shape of the decay phase even if they do adjust its levels.

\section{Derivation of Spectral Energy Transfer Functions}
\label{append:transfunc}
Spectral energy transfer analysis was first introduced in the incompressible limit by \citet{Kraichnan1967}.  Transfer analysis is a well-developed tool for studying MHD turbulence in the incompressible \citep{Debliquy2005, Verma2005} and compressible \citep{FromangPapaloizou2007a, FromangPapaloizou2007b, SimonHawleyBeckwith2009, PietarilaGraham2010} limits.  Transfer theory was outlined for compressible MHD in \citet{PietarilaGraham2010}, which was a generalization of the incompressible treatment of \citet{AlexakisMininniPouquet2005}.  Here, we expand on the transfer analysis of \citet{PietarilaGraham2010} by incorporating a decomposed velocity; thus, separating the transfer mechanisms involving the velocity field into components due to the background shear flow and turbulent motions.  This allows one to distinguish between energy transfer arising due to turbulence from that due to the background flow.

The basic philosophy behind deriving the transfer functions is to start by taking the complex conjugate of the Fourier transform of the conservation equations to obtain time-evolution equations of energy densities in Fourier space.  The Fourier transformed conservation equations are then dotted with the Fourier transform of the appropriate quantity.  The result is the time derivative of a spectral energy density being equated to many individual terms.  These terms are the transfer functions and describe energy transfer from one energy reservoir to another, mediated by a force.  In what follows, we derive the magnetic, kinetic, and internal energy transfer functions, each in turn.

The primitive form of the induction equation is,
\begin{equation}
\frac{d \mathbf{B}}{d t} = \nabla \times \left( \mathbf{v} \times \mathbf{B} \right), \label{eq:appendix_prim_induction}
\end{equation}
where $\mathbf{B}$ is the magnetic field and $\mathbf{v}$ is the fluid velocity field.  We decompose the velocity field into a turbulent velocity, $\mathbf{v}_{\rm t}$, and a shear velocity, $\mathbf{v}_{\rm sh}$, according to,
\begin{equation}
\mathbf{v} = \mathbf{v}_{\rm sh} + \mathbf{v}_{\rm t}, \label{eq:appendix_vdecomp}
\end{equation}
where,
\begin{equation}
\mathbf{v}_{\rm sh} = v_{\rm sh}(z) \widehat{\mathbf{y}} = \frac{\widehat{\mathbf{y}}}{L_{x} L_{y}} \iint v_{y} (x,y,z) dx dy.
\end{equation}
Replacing the velocity field in Equation \ref{eq:appendix_prim_induction} with the decomposed velocity defined by Equation \ref{eq:appendix_vdecomp}, taking the complex conjugate of the Fourier transform, where the Fourier transform\footnote{Fourier transforms of a quantity are denoted by a `hat', $\widehat{~~~~}$, not to be confused with hats implying unit vectors (e.g., $\widehat{\mathbf{x}}$).} of a quantity $f(\mathbf{x})$ is given by,
\begin{equation}
\widehat{F}(\mathbf{k}) = \iiint f(\mathbf{x}) e^{-i \mathbf{k} \cdot \mathbf{x}} d^{3} \mathbf{x},
\end{equation}
and dotting the result with $\widehat{\mathbf{B}}(k)$, we obtain the equation representing the transfer of magnetic energy in $k$-space,
\begin{align}
\frac{d E_{\rm M}(k)}{dt} =~&T_{\rm BBA}(k) + S_{\rm BBA}(k) + T_{\rm KBT}(k) + S_{\rm KBT}(k) + \nonumber \\
& T_{\rm KBP}(k) + D_{\rm M}(k). \label{eq:appendix_MEtransfunc}
\end{align}
The left hand side is the time-derivative of the spectral magnetic energy density, where,
\begin{equation}
E_{\rm M}(k) = \frac{1}{2} \widehat{\mathbf{B}}(k) \cdot \widehat{\mathbf{B}}^{\ast}(k).
\end{equation}
The terms on the right-hand side of Equation \ref{eq:appendix_MEtransfunc} are the magnetic energy transfer functions.  We first describe the transfer function notation and then identify each term explicitly.  Transfer functions with the notation $T_{\rm XYF}(k)$ depend only on turbulent velocities, $\mathbf{v}_{\rm t}$, those with notation $S_{\rm XYF}(k)$ depend only on the background shear velocity, $\mathbf{v}_{\rm sh}$, and those with notation $X_{\rm XYF}(k)$ have a mixed velocity dependence.  As described by \citet{PietarilaGraham2010}, the transfer function $T$, $S$, $X_{\rm XYF}(k)$ measures the net energy transfer rate from {\it all scales} of reservoir X to scale $k$ of reservoir Y, where the energy exchange is mediated by the force F.  The net energy transfer from reservoir X into reservoir Y at scale $k$ is positive (negative) for $T$, $S$, $X_{\rm XYF}(k) > 0~(< 0)$.  In other words, energy is {\it lost} by reservoir X and {\it gained} by reservoir Y at scale $k$ for  $T$, $S$, $X_{\rm XYF}(k) > 0$ and {\it vice versa} for  $T$, $S$, $X_{\rm XYF}(k) < 0$.  The available energy reservoirs are kinetic (K), magnetic (M), and internal (I).  The cascade of magnetic energy to other scales from within the magnetic energy reservoir is described by the terms,
\begin{align}
T_{\rm BBA}(k) =& - \widehat{\mathbf{B}}(k) \cdot \widehat{\left[ \left( \mathbf{v}_{\rm t} \cdot \mathbf{\nabla} \right) \mathbf{B} \right]}^{\ast}(k) \\
S_{\rm BBA}(k) =& - \widehat{\mathbf{B}}(k) \cdot \widehat{\left[ \left( \mathbf{v}_{\rm sh} \cdot \mathbf{\nabla} \right) \mathbf{B} \right]}^{\ast}(k).
\end{align}
The transfer of energy from the kinetic energy reservoir to the magnetic energy reservoir by turbulent ($T$) and shearing ($S$) fluid motions that twist and stretch field lines are,
\begin{align}
T_{\rm KBT}(k) =~& \widehat{\mathbf{B}}(k) \cdot \widehat{\left[ \left( \mathbf{B} \cdot \mathbf{\nabla} \right) \mathbf{v}_{\rm t} \right]}^{\ast}(k) \\
S_{\rm KBT}(k) =~& \widehat{\mathbf{B}}(k) \cdot \widehat{\left[ \left( \mathbf{B} \cdot \mathbf{\nabla} \right) \mathbf{v}_{\rm sh} \right]}^{\ast}(k).
\end{align}
Magnetic energy transfer from the kinetic energy reservoir by compressive motions via the magnetic pressure force is given by,
\begin{equation}
T_{\rm KBP}(k) = - \widehat{\mathbf{B}}(k) \cdot \widehat{\left[ \mathbf{B} \left( \mathbf{\nabla} \cdot \mathbf{v}_{\rm t} \right) \right]}^{\ast}(k),
\end{equation}
where there is no $S_{\rm KBP}(k)$ transfer function with a background shear velocity dependence because,
\begin{equation}
\nabla \cdot \mathbf{v}_{\rm sh} = \frac{d v_{\rm sh}(z)}{d y} = 0.
\end{equation}
Formally, the magnetic energy transfer function expression (Equation \ref{eq:appendix_MEtransfunc}) is analytically exact in the omission of the numerical magnetic energy dissipation term, $D_{\rm M}(k)$.  However, numerical schemes have dissipative effects.  Therefore, any inequality that arises from comparing the time-derivatives of spectral energy densities on the left-hand side to the sum of the transfer function terms on the right-hand side is folded into $D_{\rm M}(k)$, which is a measure of the numerical magnetic dissipation.  The other transfer function equations will have associated numerical dissipation terms as well.

The kinetic energy transfer functions are derived in a similar fashion as was done for the magnetic energy transfer functions.  Starting from the primitive form of the momentum equation,
\begin{equation}
\rho \frac{\del \mathbf{v}}{\del t} = - \rho \left( \mathbf{v} \cdot \nabla \right) \mathbf{v} - \nabla P + \left( \nabla \times \mathbf{B} \right) \times \mathbf{B}, \label{eq:appendix_prim_mom}
\end{equation}
and the conservative form of the momentum equation,
\begin{equation}
\frac{\del \left( \rho \mathbf{v} \right)}{\del t} = - \nabla \cdot \left[ \rho \mathbf{v} \mathbf{v} - \mathbf{B} \mathbf{B} + \left( P + \frac{1}{2} B^{2} \right) \mathbf{I} \right], \label{eq:appendix_cons_mom}
\end{equation}
the velocities in each of these equations are decomposed according to Equation \ref{eq:appendix_vdecomp}.  Here, $\rho$ represents the mass density, $P$ is the pressure, and $\mathbf{I}$ is the identity matrix.  The complex conjugate of the Fourier transform of Equation \ref{eq:appendix_prim_mom} is dotted with $\rho \mathbf{v}$ and the complex conjugate of the Fourier transform of Equation \ref{eq:appendix_cons_mom} is dotted with $\mathbf{v}$.  Combining these two resulting equations yields the expression representing the transfer of kinetic energy in $k$-space,
\begin{align}
\frac{d E_{\rm K}(k)}{d t} =~& T_{\rm IKC}(k) + S_{\rm IKC}(k) + T_{\rm KKA}(k) + X_{\rm KKA}(k) + \nonumber \\
& T_{\rm BKT}(k) + S_{\rm BKT}(k) + T_{\rm BKP}(k) + S_{\rm BKP}(k) + \nonumber \\
& T_{\rm KKC}(k) + S_{\rm KKC}(k) + X_{\rm KKC}(k) + D_{\rm K}(k),
\end{align}
where the spectral kinetic energy density is defined by,
\begin{equation}
E_{\rm K}(k) = \frac{1}{4} \left( \widehat{\mathbf{v}}(k) \cdot \widehat{\left[ \rho \mathbf{v} \right]} ^{\ast}(k) + \widehat{\left[ \rho \mathbf{v} \right]} (k) \cdot \widehat{\mathbf{v}}^{\ast}(k) \right).
\end{equation}
The transfer functions describing the exchange of kinetic energy from within the kinetic energy reservoir by compressible motions due to turbulence and background shear are,
\begin{align}
T_{\rm KKC}(k) =& - \frac{1}{2} \left( \widehat{\mathbf{v}_{\rm t}}(k) \cdot \widehat{\left[ \mathbf{v}_{\rm t} \left( \mathbf{\nabla} \cdot \rho \mathbf{v}_{\rm t} \right) \right]}^{\ast}(k) \right) \\
S_{\rm KKC}(k) =& - \frac{1}{2} \left( \widehat{\mathbf{v}_{\rm sh}}(k) \cdot \widehat{\left[ \mathbf{v}_{\rm sh} \left( \mathbf{\nabla} \cdot \rho \mathbf{v}_{\rm sh} \right) \right]}^{\ast}(k) \right),
\end{align}
respectively.  The corresponding cross-term transfer function is,
{\scriptsize
\begin{align}
&X_{\rm KKC}(k) = \nonumber \\
&- \frac{1}{2} \left( \widehat{\mathbf{v}_{\rm t}}(k) \cdot \widehat{\left[ \mathbf{v}_{\rm sh} \left( \mathbf{\nabla} \cdot \rho \mathbf{v}_{\rm t} \right) \right]}^{\ast}(k) + \widehat{\mathbf{v}_{\rm sh}}(k) \cdot \widehat{\left[ \mathbf{v}_{\rm t} \left( \mathbf{\nabla} \cdot \rho \mathbf{v}_{\rm t} \right) \right]}^{\ast}(k) \right) \nonumber \\
&- \frac{1}{2} \left( \widehat{\mathbf{v}_{\rm sh}}(k) \cdot \widehat{\left[ \mathbf{v}_{\rm sh} \left( \mathbf{\nabla} \cdot \rho \mathbf{v}_{\rm t} \right) \right]}^{\ast}(k) + \widehat{\mathbf{v}_{\rm t}}(k) \cdot \widehat{ \left[ \mathbf{v}_{\rm t} \left( \mathbf{\nabla} \cdot \rho \mathbf{v}_{\rm sh} \right) \right]}^{\ast}(k) \right) \nonumber \\
&- \frac{1}{2} \left( \widehat{\mathbf{v}_{\rm t}}(k) \cdot \widehat{ \left[ \mathbf{v}_{\rm sh} \left( \mathbf{\nabla} \cdot \rho \mathbf{v}_{\rm sh} \right) \right]}^{\ast}(k) + \widehat{\mathbf{v}_{\rm sh}}(k) \cdot \widehat{ \left[ \mathbf{v}_{\rm t} \left( \mathbf{\nabla} \cdot \rho \mathbf{v}_{\rm sh} \right) \right]}^{\ast}(k) \right).
\end{align}}

The transfer of energy within the kinetic energy reservoir by advection is described by the transfer functions,
{\scriptsize
\begin{align}
&T_{\rm KKA}(k) = \nonumber \\
& - \frac{1}{2} \left( \widehat{\left[ \rho \mathbf{v}_{\rm t} \right]} \cdot \widehat{\left[ \left( \mathbf{v}_{\rm t} \cdot \mathbf{\nabla} \right) \mathbf{v}_{\rm t} \right]}^{\ast}(k)+ \widehat{\mathbf{v}}_{\rm t}(k) \cdot \widehat{\left[ \left( \rho \mathbf{v}_{\rm t} \cdot \mathbf{\nabla} \right) \mathbf{v}_{\rm t} \right]}^{\ast}(k) \right) \\
&X_{\rm KKA}(k) = \nonumber \\
& - \frac{1}{2} \left( \widehat{\left[ \rho \mathbf{v}_{\rm sh} \right]}(k) \cdot \widehat{\left[ \left( \mathbf{v}_{\rm t} \cdot \mathbf{\nabla} \right) \mathbf{v}_{\rm t} \right]}^{\ast}(k) + \widehat{\mathbf{v}_{\rm sh}}(k) \cdot \widehat{\left[ \left( \mathbf{v}_{\rm t} \cdot \mathbf{\nabla} \right) \mathbf{v}_{\rm t} \right]}^{\ast}(k) \right) \nonumber \\
& - \frac{1}{2} \left( \widehat{\left[ \rho \mathbf{v}_{\rm t} \right]}(k) \cdot \widehat{\left[ \left( \mathbf{v}_{\rm sh} \cdot \mathbf{\nabla} \right) \mathbf{v}_{\rm t} \right]}^{\ast}(k) + \widehat{\mathbf{v}_{\rm t}}(k) \cdot \widehat{\left[ \left( \mathbf{v}_{\rm sh} \cdot \mathbf{\nabla} \right) \mathbf{v}_{\rm t} \right]}^{\ast}(k) \right) \nonumber \\
& - \frac{1}{2} \left(\widehat{\left[ \rho \mathbf{v}_{\rm sh} \right]}(k) \cdot \widehat{\left[ \left( \mathbf{v}_{\rm sh} \cdot \mathbf{\nabla} \right) \mathbf{v}_{\rm t} \right]}^{\ast}(k) + \widehat{\mathbf{v}_{\rm sh}}(k) \cdot \widehat{\left[ \left( \mathbf{v}_{\rm sh} \cdot \mathbf{\nabla} \right) \mathbf{v}_{\rm t} \right]}^{\ast}(k) \right) \nonumber \\
& - \frac{1}{2} \left( \widehat{\left[ \rho \mathbf{v}_{\rm sh} \right]}(k) \cdot \widehat{\left[ \left( \mathbf{v}_{\rm t} \cdot \mathbf{\nabla} \right) \mathbf{v}_{\rm sh} \right]}^{\ast}(k) + \widehat{\mathbf{v}_{\rm sh}}(k) \cdot \widehat{\left[ \left( \rho \mathbf{v}_{\rm t} \cdot \mathbf{\nabla} \right) \mathbf{v}_{\rm sh} \right]}^{\ast}(k) \right) \nonumber \\
& - \frac{1}{2} \left( \widehat{\left[ \rho \mathbf{v}_{\rm t} \right]}(k) \cdot \widehat{\left[ \left( \mathbf{v}_{\rm t} \cdot \mathbf{\nabla} \right) \mathbf{v}_{\rm sh} \right]}^{\ast}(k) + \widehat{\mathbf{v}_{\rm t}}(k) \cdot \widehat{\left[ \left( \rho \mathbf{v}_{\rm t} \cdot \mathbf{\nabla} \right) \mathbf{v}_{\rm sh} \right]}^{\ast}(k)\right).
\end{align}}

Generally speaking, the transfer functions involving mixed velocity terms, which are denoted by $X_{\rm XYF}(k)$, are not intuitively graspable.  However, these cross-terms only appear for the transfer functions describing the energy cascade within the kinetic energy reservoir by compressive motions and advection.  Energy transferred from the magnetic energy reservoir to the kinetic energy reservoir by fluid motions via the magnetic tension force are,
{\scriptsize
\begin{align}
T_{\rm BKT}(k) =~& \frac{1}{2} \left( \widehat{\left[ \rho \mathbf{v}_{\rm t} \right]}(k) \cdot \widehat{\left[ \frac{1}{\rho} \left( \mathbf{B} \cdot \mathbf{\nabla} \right) \mathbf{B} \right]}^{\ast}(k) + \widehat{\mathbf{v}_{\rm t}}(k) \cdot \widehat{\left[ \left( \mathbf{B} \cdot \mathbf{\nabla} \right) \mathbf{B} \right]}^{\ast}(k) \right) \\
S_{\rm BKT}(k) =~& \frac{1}{2} \left( \widehat{\left[ \rho \mathbf{v}_{\rm sh} \right]}(k) \cdot \widehat{\left[ \frac{1}{\rho} \left( \mathbf{B} \cdot \mathbf{\nabla} \right) \mathbf{B} \right]}^{\ast}(k) + \widehat{\mathbf{v}_{\rm sh}} \cdot \widehat{\left[ \left( \mathbf{B} \cdot \mathbf{\nabla} \right) \mathbf{B} \right]}^{\ast}(k) \right),
\end{align}}
and by compressive turbulent fluid motions via the magnetic pressure force are,
{\scriptsize
\begin{align}
T_{\rm BKP}(k) =& - \frac{1}{2} \left( \widehat{\left[ \rho \mathbf{v}_{\rm t} \right]}(k) \cdot \widehat{\left[ \frac{1}{2 \rho} \mathbf{\nabla} B^{2} \right]}^{\ast}(k) + \widehat{\mathbf{v}_{\rm t}} \cdot \widehat{\left[ \frac{1}{2} \mathbf{\nabla} B^{2} \right]}^{\ast}(k) \right) \\
S_{\rm BKP}(k) =& - \frac{1}{2} \left( \widehat{\left[ \rho \mathbf{v}_{\rm sh} \right]}(k) \cdot \widehat{\left[ \frac{1}{2\rho} \mathbf{\nabla} B^{2} \right]}^{\ast}(k) + \widehat{\mathbf{v}_{\rm sh}} \cdot \widehat{\left[ \frac{1}{2} \mathbf{\nabla} B^{2} \right]}^{\ast}(k) \right).
\end{align}}

Finally, the transfer functions describing energy exchange from the internal energy reservoir to the kinetic energy reservoir by compressive motions are,
{\scriptsize
\begin{align}
T_{\rm IKC}(k) =& - \frac{1}{2} \left( \widehat{\left[ \rho \mathbf{v}_{\rm t} \right]}(k) \cdot \widehat{\left[ \frac{1}{\rho} \mathbf{\nabla} P \right]}^{\ast}(k) + \widehat{\mathbf{v}_{\rm t}} \cdot \widehat{\left[ \mathbf{\nabla} P \right]}^{\ast}(k) \right) \\
S_{\rm IKC}(k) =& - \frac{1}{2} \left( \widehat{\left[ \rho \mathbf{v}_{\rm sh} \right]}(k) \cdot \widehat{\left[ \frac{1}{\rho} \mathbf{\nabla} P \right]}^{\ast}(k) + \widehat{\mathbf{v}_{\rm sh}} \cdot \widehat{\left[ \mathbf{\nabla} P \right]}^{\ast}(k) \right).
\end{align}}

Again, following the same procedure as for deriving the magnetic and kinetic energy transfer functions, we start with the internal energy equation,
\begin{equation}
\frac{\del P}{\del t} = - \mathbf{v} \cdot \nabla P + \gamma P \left( \nabla \cdot \mathbf{v} \right), \label{eq:appendix_prim_internal}
\end{equation}
where $\gamma$ is the adiabatic index.  Decomposing the velocity in this equation according to Equation \ref{eq:appendix_vdecomp} and multiplying the complex conjugate of the Fourier transform by $\widehat{P}(k)$, the equation describing the transfer of internal energy in $k$-space becomes,
\begin{equation}
\frac{d E_{\rm I}(k)}{d t} = T_{\rm KIA}(k) + S_{\rm KIA}(k) + T_{\rm KIC}(k) + D_{\rm I}(k).
\end{equation}
The internal energy density in $k$-space is defined as,
\begin{equation}
E_{\rm I}(k) = \frac{\widehat{P}(k)}{\gamma -1},
\end{equation}
and the transfer functions associated with energy exchange from the kinetic energy reservoir to the internal energy reservoir by advection and compressive motions are,
\begin{align}
T_{\rm KIA}(k) =& - \frac{1}{\gamma - 1} \widehat{\sqrt{P}}(k) \widehat{\left[ \frac{1}{\sqrt{P}} \left( \mathbf{v}_{\rm t} \cdot \mathbf{\nabla} \right) P \right]}^{\ast}(k) \\
S_{\rm KIA}(k) =& - \frac{1}{\gamma - 1} \widehat{\sqrt{P}}(k) \widehat{\left[ \frac{1}{\sqrt{P}} \left( \mathbf{v}_{\rm sh} \cdot \mathbf{\nabla} \right) P \right]}^{\ast}(k) \\
T_{\rm KIC}(k) =& \frac{\gamma}{\gamma - 1} \widehat{\sqrt{P}}(k) \widehat{\left[ \frac{1}{\sqrt{P}} P \left( \mathbf{\nabla} \cdot \mathbf{v}_{\rm t} \right) \right]}^{\ast}(k).
\end{align}

The transfer function analysis presented above can be extended to the case of a viscous and resistive fluid.  We derive these additional dissipation transfer function terms following the procedure outlined in \citet{FromangPapaloizou2007a} and \citet{SimonHawleyBeckwith2009}.  Turning our attention first to the induction equation, the inclusion of Ohmic resistivity introduces the term, $\eta \nabla^{2} \mathbf{B}$, to the right-hand side of Equation \ref{eq:appendix_prim_induction}, where $\eta$ is the resistivity.  Taking the complex conjugate of the Fourier transform for this Ohmic dissipation term and dotting it with $\widehat{\mathbf{B}}(k)$ yields,
\begin{equation}
T_{\eta}(k) = \eta \left( \widehat{\mathbf{B}}(k) \cdot \widehat{\left[ \nabla^{2} \mathbf{B} \right]}^{\ast}(k) \right)
\end{equation}

Incorporating viscosity would add the terms, $(\nabla \cdot \bm{\tau}) / \rho$, and, $(\nabla \cdot \bm{\tau})$, to the right-hand sides of Equations \ref{eq:appendix_prim_mom} and \ref{eq:appendix_cons_mom}, respectively, where the stress tensor for an isotropic fluid is,
\begin{equation}
\tau_{ij} = 2 \mu \left( e_{ij} - \frac{1}{3} \left( \nabla \cdot \mathbf{v} \right) \delta_{ij} \right),
\end{equation}
where $\mu$ is the dynamic viscosity, $\delta_{ij}$ is the Kronecker delta function, and the strain rate tensor is given by,
\begin{equation}
e_{ij} = \frac{1}{2} \left[ \left( \nabla \mathbf{v} \right) + \left( \nabla \mathbf{v} \right)^{\rm T} \right].
\end{equation}
One can show that,
\begin{align}
\nabla \cdot \bm{\tau} =~& \mu \nabla \cdot \left( \left[ \left( \nabla \mathbf{v} \right) + \left( \nabla \mathbf{v} \right)^{\rm T} \right] - \frac{2}{3} \left( \nabla \cdot \mathbf{v} \right) \delta_{ij} \right) \nonumber \\
=~& \mu \left( \nabla^{2} \mathbf{v} + \frac{1}{3} \nabla \left( \nabla \cdot \mathbf{v} \right) \right).
\end{align}
The dynamic viscosity is related to the kinematic viscosity, $\nu$, by $\nu = \mu / \rho$.  Taking the complex conjugate of the Fourier transform of the viscous term and dotting the conservative form with $\widehat{\mathbf{v}}(k)$ and the primitive form with $\widehat{[ \rho \mathbf{v}]}(k)$ gives the transfer function describing the viscous dissipation,
\begin{equation}
T_{\nu}(k) = \frac{\nu}{2} \left( \widehat{\left[ \rho \mathbf{v} \right]} \cdot \widehat{\left[ \frac{1}{\rho} \left( \nabla \cdot \bm{\tau} \right) \right]}^{\ast}(k) + \widehat{\mathbf{v}} \cdot \widehat{\left( \nabla \cdot \bm{\tau} \right)}^{\ast}(k) \right).
\end{equation}

Note that due to the non-linear nature of the resistive and viscous terms, we choose not to decompose the velocity field in the definitions of $T_{\eta}(k)$ and $T_{\nu}(k)$ in order to simplify their interpretations.

\end{document}